\documentclass{aa}  

\usepackage{graphicx,color}
\usepackage{txfonts}
\usepackage{xspace}
\usepackage[usenames,dvipsnames,table]{xcolor} 

\usepackage{colortbl}       
\usepackage{booktabs}       
\usepackage{array}          
\usepackage[normalem]{ulem}

\usepackage{fourier}
\DeclareMathAlphabet{\mathcal}{OMS}{cmsy}{m}{n}
\SetMathAlphabet{\mathcal}{bold}{OMS}{cmsy}{b}{n}

\usepackage[colorlinks,bookmarks]{hyperref}
\definecolor{linkblue}{rgb}{0,0,0.8}
\definecolor{linkgreen}{rgb}{0,0.5,0}
\hypersetup{linkcolor=linkblue, citecolor=linkgreen, urlcolor=linkblue}

\definecolor{valecol}{rgb}{0,0.5, 1.}

\definecolor{ttcol}{rgb}{1.,0.5, 0.}

\makeatletter
\newcommand{\vast}{\bBigg@{4}}
\newcommand{\Vast}{\bBigg@{5}}
\makeatother

\newcommand{\DM}{D_\mathrm{M}}
\def\d{{\rm d}}

\newcommand{\Mhalo}{{\ifmmode{M_{\rm halo}}\else{$M_{\rm halo}$}\fi}}

\graphicspath{{./figs/}}

\begin{document} 

\title{Accurate cosmological emulator for the probability distribution function of gravitational lensing of point sources}

\titlerunning{Lensing PDF emulator}
\authorrunning{Turker et al.}

\author{
Tunç Turker\inst{\ref{oats},\ref{ppgcosmo}}\thanks{\email{tuncturker@gmail.com}}
\and
Valerio Marra\inst{\ref{ufes},\ref{oats},\ref{ifpu}}
\and
Tiago Castro\inst{\ref{oats},\ref{ifpu},\ref{infnTS}, \ref{icsc}}
\and
Miguel Quartin\inst{\ref{ppgcosmo},\ref{cbpf},\ref{ov}} 
\and
Stefano Borgani\inst{\ref{oats},\ref{ifpu},\ref{infnTS},\ref{units}, \ref{icsc}}
}

\institute{
INAF -- Osservatorio Astronomico di Trieste, via Tiepolo 11, 34131 Trieste, Italy \label{oats}
\and
PPGCosmo, Universidade Federal do Espírito Santo, 29075-910, Vitória, ES, Brazil \label{ppgcosmo}
\and
Departamento de Física, Universidade Federal do Espírito Santo, 29075-910, Vitória, ES, Brazil \label{ufes}
\and
IFPU -- Institute for Fundamental Physics of the Universe, via Beirut 2, 34151, Trieste, Italy 
\label{ifpu}
\and
INFN -- Sezione di Trieste, Via Valerio 2, 34127, Trieste, Italy \label{infnTS}
\and
Dipartimento di Fisica, Sezione di Astronomia, Università di Trieste, 34143, Trieste, Italy \label{units}
\and
ICSC – Centro Nazionale di Ricerca in High Performance Computing, Big Data e Quantum Computing, Bologna, Italy \label{icsc}
\and
Centro Brasileiro de Pesquisas Físicas, 22290-180, Rio de Janeiro, Brazil \label{cbpf}
\and
Observatório do Valongo, Universidade Federal do Rio de Janeiro, 20080-090, Rio de Janeiro, RJ, Brazil \label{ov}
}

\date{Received \today\ / Accepted \today}

\abstract
{}
{We develop an accurate and computationally efficient emulator to model the gravitational lensing magnification probability distribution function (PDF), enabling robust cosmological inference of point sources such as supernovae and gravitational-wave observations.}
{We construct a pipeline utilizing cosmological $N$-body simulations, creating past light cones to compute convergence and shear maps. Principal Component Analysis (PCA) is employed for dimensionality reduction, followed by an eXtreme Gradient Boosting (XGBoost) machine learning model to interpolate magnification PDFs across a broad cosmological parameter space ($\Omega_{\rm m}$, $\sigma_8$, $w$, $h$) and redshift range ($0.2 \leq z \leq 6$). We identify the optimal number of PCA components to balance accuracy and stability.}
{Our emulator, publicly released as \href{https://github.com/Turkero/ACE-Lensing}{\texttt{ace\_lensing}}, accurately reproduces lensing PDFs with a median Kullback–Leibler divergence of $0.007$.
Validation on the test set confirmed that the model reliably reproduces the detailed shapes and statistical properties of the PDFs across the explored parameter range, showing no significant degradation for specific parameter combinations or redshifts.
Future work will focus on incorporating baryonic physics through hydrodynamical simulations and expanding the training set to further enhance model accuracy and generalizability.}
{}

\keywords{gravitational lensing: weak -- large-scale structure of Universe -- methods: numerical -- methods: statistical}

\maketitle

\section{Introduction}

Gravitational lensing magnification, caused by the intervening large-scale structure, encodes valuable cosmological information by altering the observed brightness and angular sizes of distant sources. This effect both impacts and enhances cosmological and astrophysical inferences drawn from, for example, supernovae and gravitational-wave observations.

In this work, we focus on the 1-point statistics of lensing magnification, which has emerged as a valuable complement to traditional analyses based on 2-point statistics \citep[see, e.g.,][]{Friedrich:2025nao}. Unlike the latter, which only exploits the variance and covariance of observables, the 1-point statistics captures the full non-Gaussian structure of the magnification probability distribution function (PDF)~\citep[see, e.g.,][]{Patton:2016umg,Liu:2018dsw,Uhlemann:2019gni,Uhlemann:2022znd}.
Accurate modeling of the magnification PDF is essential for unbiased cosmological inference from magnification data, enabling not only accurate corrections to the observed luminosity distances of sources, which can otherwise bias supernovae~\citep{Pierel:2020tav} or gravitational wave distances~\citep{Shan:2020esq,Canevarolo:2023dkh,Ferri:2024amc}, but also the extraction of cosmological information directly from the measured PDF, as in the \emph{method of the moments} \cite[MeMo,][]{Quartin:2013moa}.
However, existing analytic approximations and simplified methods lack the precision required to capture nonlinear structures, particularly in the high-magnification regime.

Fully nonlinear cosmological simulations have become essential tools for obtaining accurate lensing PDFs, as demonstrated, for example, by \citet{Alfradique:2024fkb}. Simplified semi-analytic approaches suffer from biases in lensing moments due to limitations in capturing small-scale nonlinearities and halo internal structures, ultimately impacting cosmological inference. Fully analytical approaches also exist, but are limited to the weak-lensing limit and to the convergence PDF~\citep{Thiele:2020rig}.
Hydrodynamical simulations incorporating baryonic physics significantly alter the lensing PDF at high magnifications \citep{Castro:2017tbn}. Nevertheless, due to computational constraints, dark-matter-only (DMO) simulations remain widely used. While this approximation introduces some inaccuracies, they are expected to be minor in most practical scenarios, such as applications to supernova and gravitational-wave observations.

In this work, we construct a comprehensive pipeline to generate accurate lensing PDFs using a suite of 70 $N$-body simulations spanning broad ranges in cosmological parameters and redshift. The pipeline, summarized in Appendix~\ref{ap:pipeline}, includes the construction of past light cones (PLCs), computation of convergence and shear maps, Principal Component Analysis (PCA) for dimensionality reduction, and machine learning (ML) modeling using 
the eXtreme Gradient Boosting algorithm\footnote{\url{https://github.com/dmlc/xgboost}} \citep[\texttt{XGBoost},][]{Chen:2016btl}. The final product is an efficient emulator capable of interpolating magnification PDFs across the studied parameter space, providing a powerful and computationally inexpensive tool for cosmological analyses.
The \texttt{ace\_lensing} emulator is publicly available.\footnote{\url{https://github.com/Turkero/ACE-Lensing}}

This work is organized as follows. In Section~\ref{sec:PDF}, we describe the part of the pipeline responsible for generating the lensing PDF, from numerical simulations to dataset construction, including numerical tests. Section~\ref{sec:features} addresses dimensionality reduction and feature extraction, while Section~\ref{sec:ML} presents the machine learning model employed. The emulator’s performance is evaluated in Section~\ref{sec:perf}, and conclusions are drawn in Section~\ref{conclusions}.

\section{PDF generation}
\label{sec:PDF}

\subsection{Simulations}
\label{sec:sim}

\citet{Alfradique:2024fkb} concluded that fully nonlinear cosmological simulations are necessary to obtain an accurate lensing PDF. Approximate methods such as \texttt{PINOCCHIO} \citep{Monaco:2001jg,Taffoni:2001jh,Munari:2016aut} and \texttt{turboGL} \citep{Kainulainen2009,Kainulainen2011a,Kainulainen2011b} currently fail to model lensing accurately at large values of magnification ($\mu \gtrsim 1.5$). This is due to their reduced precision in representing small-scale nonlinear matter fields, stemming from oversimplified assumptions about halo internal structure and a reliance on perturbation theory. These limitations introduce biases in the lensing moments, which would in turn propagate as biases in cosmological parameter estimates when using, for example, the \emph{method of the moments}~\citep{Quartin:2013moa,Castro:2014oja}.

\citet{Castro:2017tbn} showed that baryonic feedback significantly affects the lensing PDF at $\mu \gtrsim 3$, and that full-physics (FP) hydrodynamical simulations—including the key astrophysical processes governing galaxy formation and evolution—are recommended for fully accurate modeling. 
Here, however, we adopt dark-matter-only (DMO) simulations due to the high computational cost of FP simulations, deferring to future work the task of calibrating the lensing PDF emulator using a suitable suite of hydrodynamical simulations. We expect the bias introduced by this approximation to be small for applications to supernova and gravitational-wave observations. This is because full-physics and dark-matter-only simulations differ most strongly in the high-magnification tail of the distribution—corresponding to the multiple-imaging regime—where local discrepancies have little impact on global statistical properties.
To quantify this, \citet{Alfradique:2024fkb} introduced the $\mathcal{L}_1$ norm, defined as the weighted relative difference between a surrogate and a reference PDF. For example, a local discrepancy of 50\% in the PDF's high magnification tail corresponds to a weighted relative difference of only 2\% in the magnification statistics.

\begin{table}
\caption{Cosmological parameters and simulation setup.}
\label{tab:sims}
\centering
\setlength{\tabcolsep}{10pt}
\renewcommand{\arraystretch}{1.2}
\begin{tabular}{lc}
\toprule
\multicolumn{2}{c}{Cosmological parameters} \\
\midrule
$\Omega_{\rm m}$ & $[0.2, 0.4]$ \\
$\sigma_8$       & $[0.7, 0.9]$ \\
$w$              & $[-1.5, -0.7]$ \\
$h$              & $[0.6, 0.8]$ \\
\midrule
\multicolumn{2}{c}{Simulation parameters} \\
\midrule
Box size $L$              & $250\,\text{Mpc}\, h^{-1}$ \\
Number of particles $N$   & $4\times640^3$ \\
Particle-Mesh grid    & $2048^3$ \\
Initial redshift      & $49$ \\
Softening length      & $6\,\text{kpc}\, h^{-1}$ (comoving) \\
Particle mass         & $4.1\,\Omega_{\rm m} \times 10^9\,M_\odot\, h^{-1}$ \\
\bottomrule
\end{tabular}
\end{table}

For our DMO simulations, we employed the Tree-PM gravity solver of the \texttt{Opengadget3} code \citep{Dolag:2026}.
We adopted a box size of $250\,\mathrm{Mpc}\,h^{-1}$, which is sufficiently large to suppress significant cosmic variance and mode-coupling effects \citep{Castro:2017tbn} and, with a computationally inexpensive setup of $4\times640^3$ particles,\footnote{FCC lattice \texttt{MonofonIC} configuration \citep{Michaux:2020yis}.} yields a particle mass resolution of $\sim 10^9\,M_\odot\,h^{-1}$. This resolution ensures an effective modeling of small-scale nonlinear structures relevant for lensing \citep{Castro:2017tbn}.
The chosen resolution guarantees an accurate reproduction of the matter clustering down to the scales where baryonic feedback becomes relevant. The inclusion of baryonic effects will be addressed in a forthcoming work.

\begin{figure}
\centering 
\includegraphics[width= \columnwidth]{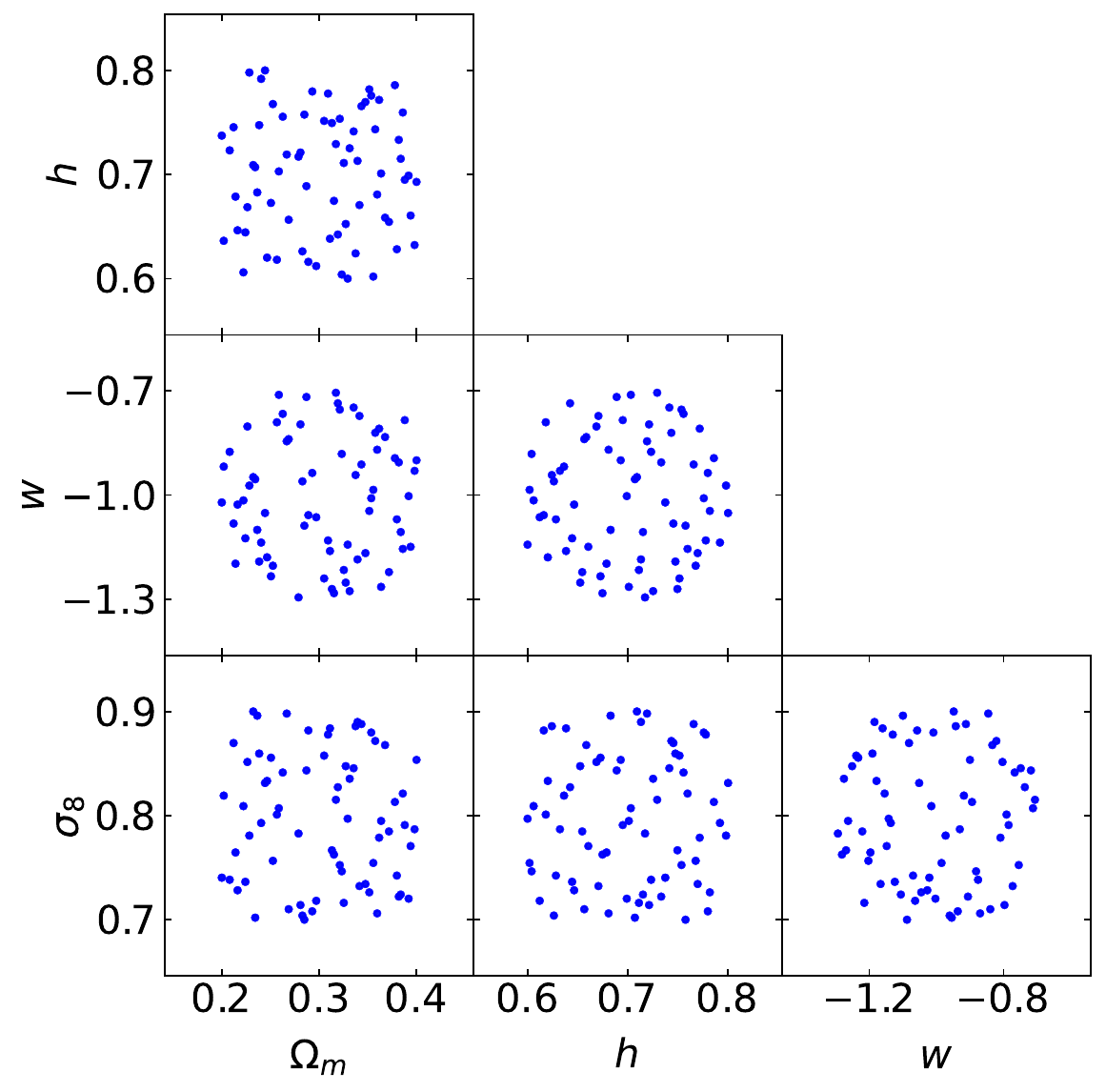}
\caption{Distribution of the values of the cosmological parameters for the 70 cosmologies here employed to build the lensing PDF emulator. These values were generated via Latin hypercube sampling.}
\label{fig:hypercube}
\end{figure}

The initial condition snapshots were generated using \texttt{MonofonIC}\footnote{\url{https://bitbucket.org/ohahn/monofonic}}, using third-order Lagrangian perturbation theory, and subsequently provided to Opengadget3 at a redshift of  $z = 49$. To reduce the effect of cosmic variance we fixed the initial power spectrum amplitudes \citep{Angulo:2016hjd}. We conducted a total of 70 simulations with the parameters reported in Table~\ref{tab:sims}. In all simulations, the spatial-curvature density parameter was set to 0, and the dark energy equation of state was assumed to be non-dynamical, with no time evolution. We adopted a fixed baryon density parameter of 0.0494 and a primordial scalar spectral index of 0.9661, defined at the pivot scale 0.05 $\mathrm{Mpc}^{-1}$.
Neutrinos were modeled as 3 species, with 2 assumed to be massless and one massive species with mass 0.06 $\mathrm{eV}$. For each simulation, we stored approximately 30 snapshots covering the redshift range from 10 to 0, yielding a total of about 2000 snapshots and, as explained below, corresponding lensing PDFs. To systematically sample the parameter space, we employed the Latin Hypercube Sampling (LHS) strategy.
We employ an LHS technique based on~\cite{DeRose:2018xdj}, but instead of maximizing the distance between the points based on their 2D projections, we maximize their full 4D Euclidean distance, which results in a better 4D point spread. This ensures an efficient and representative coverage of the multidimensional parameter space for our simulations. This parameter space is visualized in the Figure~\ref{fig:hypercube}.

\begin{figure}
\centering 
\includegraphics[width= \columnwidth]{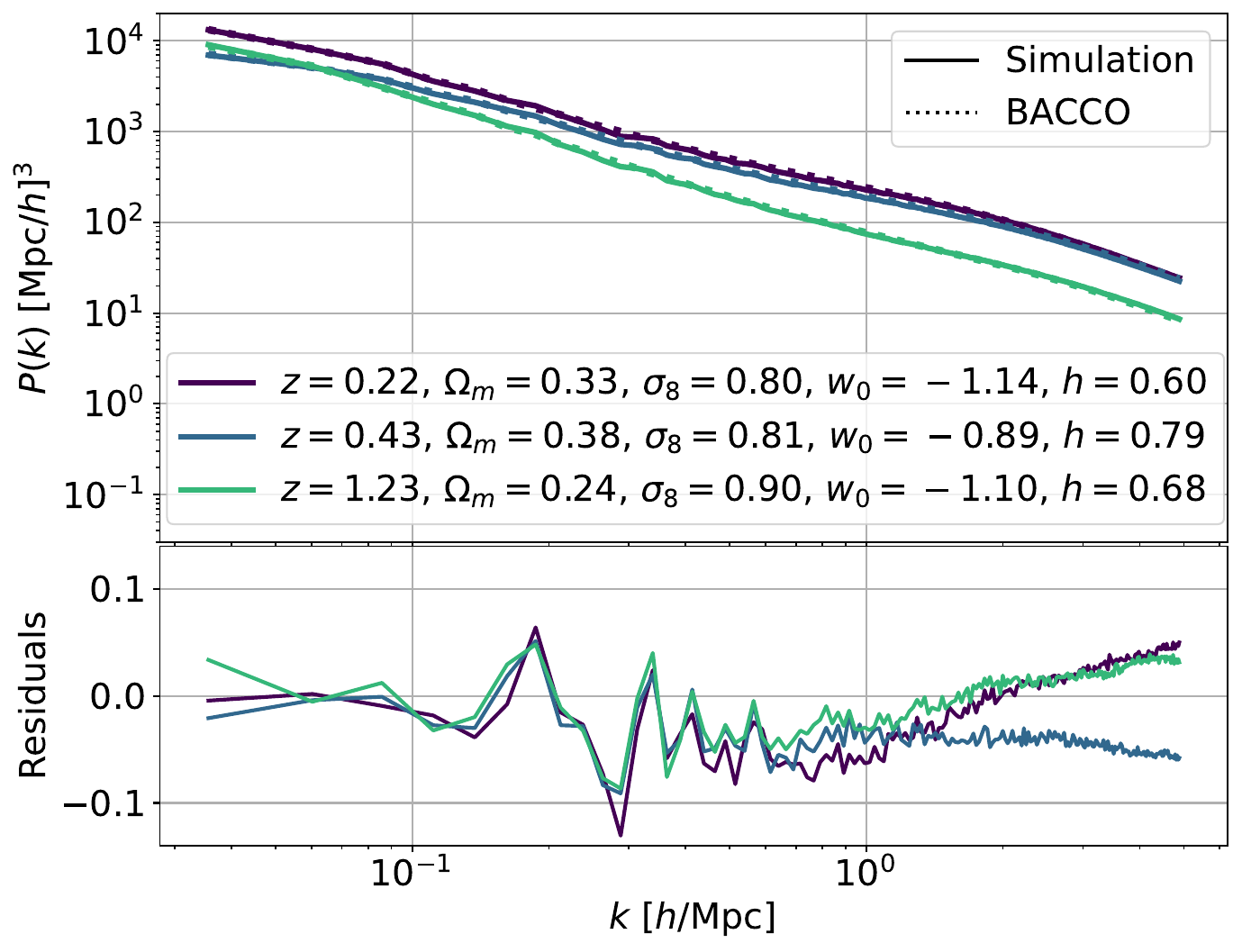}
\caption{Comparison between the matter power spectrum measured from the simulation snapshots and the nonlinear prediction from \texttt{baccoemu}.}
\label{fig:pk}
\end{figure}

Figure~\ref{fig:pk} compares the matter power spectrum measured from our simulations with the nonlinear prediction of \texttt{baccoemu}\footnote{\url{https://baccoemu.readthedocs.io}} \citep{Angulo:2020vky}, which is calibrated on $N$-body runs with larger volumes and higher mass resolution. The simulation results closely follow the emulator prediction.

\begin{figure}
\centering 
\includegraphics[width= \columnwidth]{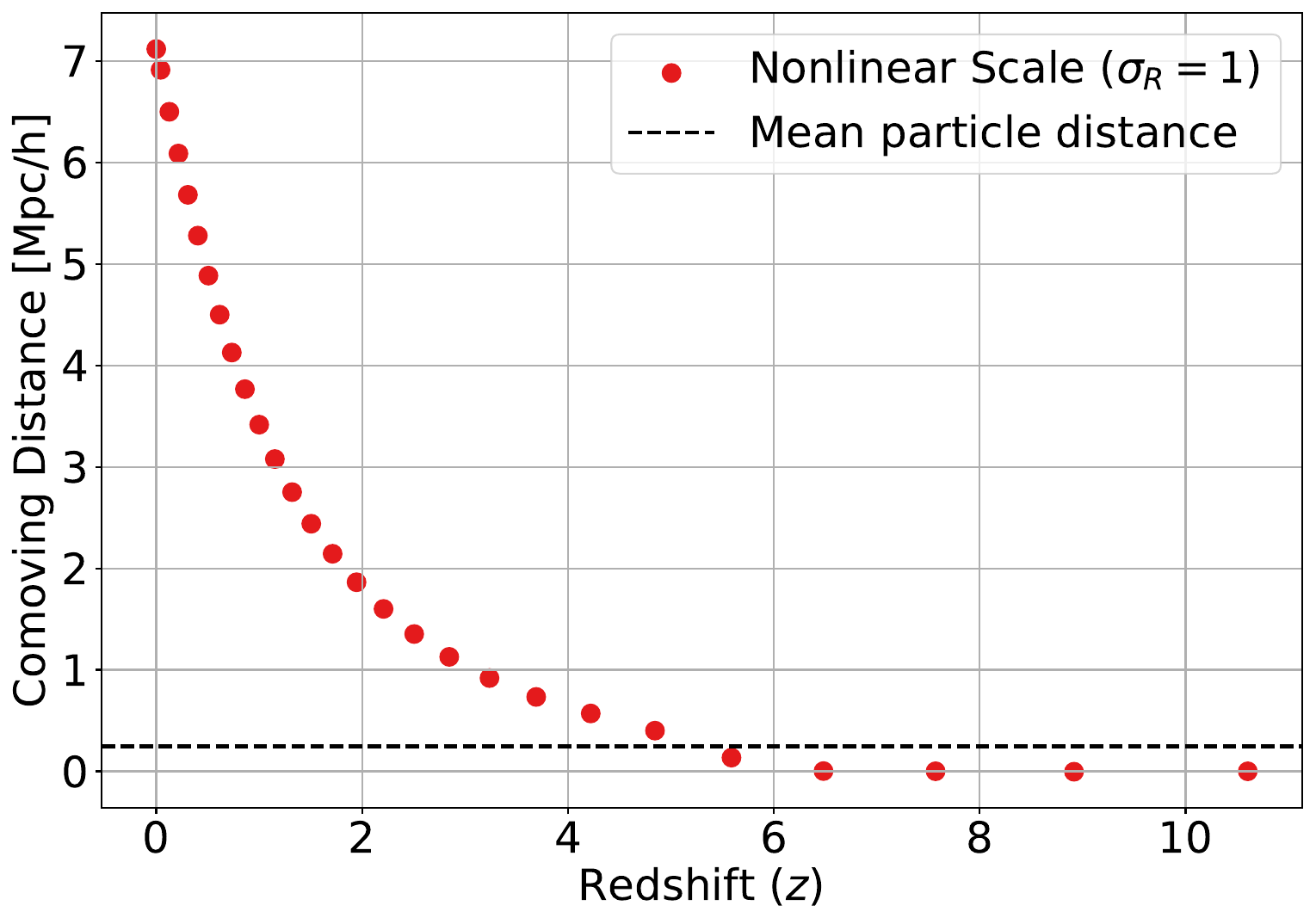}
\caption{Redshift dependence of resolvable matter fluctuations. The dots show the scale at which density fluctuations reach unity ($\sigma_R = 1$), and the black dashed line indicates the simulation’s mean particle separation. Regions below the dashed line are not reliably captured by the simulation due to resolution limits.}
\label{fig:density_fluc_scale}
\end{figure}

To assess the limitations of our simulations, we examined the scale at which matter perturbations enter the nonlinear regime and compared it to the mean inter-particle distance, $L/N^{1/3}$. The nonlinear scale is defined as the scale $R$ at which the variance of matter fluctuations satisfies $\sigma_R^2 = 1$, using a top-hat window function. 
For the computation of $\sigma_R$, we use the actual matter power spectrum measured from the snapshots of simulation C10, which has parameters $\Omega_{\rm m}=0.331$, $\sigma_8=0.835$, $w=-1.28$, and $h=0.725$.
Figure~\ref{fig:density_fluc_scale} shows that for $z \gtrsim 6$ the nonlinear scale drops below the interparticle distance, indicating that nonlinear clustering is suppressed. We thus restrict our PLCs to the redshift interval $0.2 \leq z \leq 6$ to guarantee that the analysis remains within the regime of sufficient simulation resolution.\footnote{Since lensing efficiency approaches zero at the source position, contributions from the matter field at $z \approx 6$ have negligible impact on our results.} Consequently, we utilize approximately 20 out of the 30 available snapshots for each simulation.\footnote{The precise number depends on the cosmological parameters specific to each simulation.}

\subsection{PLC construction technique}

To construct past light cones (PLCs) for each simulation, we sequentially stack the snapshots, applying random rotations and translations before placing them along the line of sight to minimize correlations and boundary effects.

Subsequently, the lensing PDF is computed by tracing light rays through the PLC. To this end, we divide the resulting image into pixels, the size of which introduces an additional resolution scale, distinct from the intrinsic softening scale and particle mass of the simulation. Each snapshot is projected onto a single lens plane that is perpendicular to the observer’s line of sight and passes through the center of the randomized simulation box. Consequently, the total number of lens planes coincides with the number of snapshots, approximately 30.

For the construction of past light cones (PLCs), we employed two complementary computational approaches, namely \texttt{SLICER}\footnote{\url{https://github.com/TiagoBsCastro/SLICER}} and \texttt{PAINLESS}.\footnote{\texttt{PAINLESS} is a sub-routine of \texttt{SLICER}.}
These methods differ primarily in how they handle spatial sampling across redshift.

The \texttt{SLICER} method constructs PLCs by tracing light rays along the observer’s past light cone, maintaining a constant angular resolution $\theta_{\text{grid}}$ at all redshifts. This technique closely mimics observational conditions—such as those of wide-field imaging surveys—by preserving a fixed angular pixel scale, although the corresponding physical scale increases with distance. By construction, this approach samples only portions of the simulation boxes at each redshift, necessitating the generation of multiple PLC realizations (with varying random seeds) to fully exploit the simulation boxes.

Conversely, the \texttt{PAINLESS} method constructs PLCs by stacking slices from successive simulation snapshots, projected onto planes with a fixed spatial (comoving) grid size $r_{\text{grid}}$. This ensures a consistent physical resolution across redshift and utilizes all simulation particles, thus simplifying lensing calculations and eliminating the need for multiple PLC realizations. Although this method does not trace particles along the true past light cone, it provides a computationally efficient approximation well suited for theoretical analyses involving one-point statistics, such as the PDF of convergence $\kappa$ or shear $\gamma$. However, due to its inherent loss of angular coherence across different redshifts, it is not appropriate for analyses based on two-point statistics, such as the angular power spectrum.

In Appendix~\ref{ap:slicingpain} we compare the \texttt{SLICER} and \texttt{PAINLESS} methods, finding a good agreement between both methods.
We choose \texttt{PAINLESS} for the analysis presented below, as it offers better control over spatial sampling. In \texttt{SLICER}, the angular resolution is fixed, so the comoving pixel size decreases toward low redshift. As a result, each pixel contains fewer particles, enhancing particle discreteness effects in the projected maps (since shot noise scales as $1/n$, with $n$ the number of particles per pixel) that are not associated with physical lensing. By adopting a fixed comoving grid at all redshifts, \texttt{PAINLESS} keeps $n$ more uniform across the light cone, giving better control over these numerical fluctuations. While the \texttt{PAINLESS} approach is less suited for mimicking observational conditions, it is more appropriate for theoretical studies in cosmological lensing, where full resolution control is critical.

\subsection{Lensing maps}

We constructed convergence and shear maps by implementing the gravitational lensing formalism \citep{Bartelmann:1999yn}. The PLC construction yields maps of the surface mass density $\Sigma_i$ on successive lens planes, each representing the projected mass distribution of the corresponding simulation snapshot.

First, we computed the effective comoving lens distance $D_l$ and its corresponding redshift $z_l$ by performing a distance-weighted average over the redshift interval spanned by each lens slice:
\begin{equation}
z_l = \frac{\int_{z_{\rm low}}^{z_{\rm up}} z \, \DM(z) \, dz}{\int_{z_{\rm low}}^{z_{\rm up}} \DM(z) \, dz}\,, \qquad 
D_l= \DM(z_l) \,,
\end{equation}
where $z_{\rm low}$ and $z_{\rm up}$ delimit the redshift boundaries of the box within the PLC, $\DM (z) = \int_0^z c\, \d z^\prime/H(z^\prime)$ is the transverse comoving distance, and $c$ is the speed of light.

Next, for a given source redshift $z_s$, we identified all lens planes at redshifts $z_i < z_s$ and processed their associated projected mass density maps $\Sigma_i$ to obtain the convergence map~$\kappa(\mathbf{x})$:
\begin{equation}
\kappa(\mathbf{x}) = \frac{4\pi G}{c^2} \sum_i W(z_i, z_s) \, \left[ \Sigma_i(\mathbf{x}) - \langle \Sigma_i \rangle \right] \, (1+z_i)^2 \, \frac{G(z_i)}{G(z_{\text{snap}})} \,,
\end{equation}
where $W(z_i, z_s)=D_{l_i} D_{l_i s}/D_s$ is the lensing efficiency kernel, computed from the comoving distances between the observer, lens, and source, and the factor $(1+z_i)^2$ accounts for the proper-to-comoving scaling in the Poisson equation. The last term corrects for the mismatch between the lens-plane redshift and the snapshot redshift via the ratio of the linear growth functions $G$. This correction is small since the snapshots were saved at redshifts very close to those of the lens planes.

Finally, we computed the lensing potential $\phi$ from the convergence maps by solving the Poisson equation in Fourier space:
\begin{equation}
\tilde{\phi}(\mathbf{k}) = -\frac{2}{|\mathbf{k}|^2} \tilde{\kappa}(\mathbf{k})\,,
\end{equation}
and transformed it back to real space.
This procedure takes into account the correct periodic boundary conditions as \texttt{PAINLESS} uses the full box.
The shear fields were derived as second-order derivatives of the potential:
\begin{align}
\gamma_1 = \frac{1}{2}(\partial_{xx} \phi - \partial_{yy} \phi) \,, \qquad
\gamma_2 = \partial_{xy} \phi \,.
\end{align}

These maps are crucial for modeling gravitational lensing by large-scale structure, encapsulating how matter distributions influence the propagation of light. The convergence field ($\kappa$) characterizes isotropic magnification, while the shear field ($\gamma$) quantifies the anisotropic distortions arising from differential gravitational deflection. From these convergence and shear maps, we computed magnification maps ($\mu$), which serve as the foundation for our emulator framework. The magnification factor, describing the amplification or de-amplification of background sources due to gravitational lensing, is computed through:
\begin{gather}
    \mu = \frac{1}{ (1 - \kappa)^2 - \gamma^2} \,,
\end{gather}
which follows directly from the Jacobian determinant of the lensing transformation and quantifies the total modification in the apparent brightness and angular size of background sources.
Since magnification represents a flux ratio, which is intrinsically positive, we consider the absolute value \(|\mu|\).

The magnification field is especially relevant for weak-lensing analyses, as it affects the observed luminosity of distant point-like objects (e.g., supernovae) and modifies source number counts at various flux thresholds. Furthermore, the statistical distribution of magnification encodes essential cosmological information regarding matter density fluctuations and the evolution of large-scale structure.

After generating the maps, we computed the magnification probability distribution function (PDF), which captures the statistical properties of the magnification field across various cosmological models and source redshifts. Our analysis spans 1447 distinct configurations of cosmological parameters and redshift values ($\sim 70 \times 20$)\footnote{As discussed in Section~\ref{sec:sim} and Figure~\ref{fig:density_fluc_scale}, we restrict the analysis to the PDFs relative to $z < 6$.}, enabling a comprehensive investigation of lensing-induced magnification effects throughout the parameter space. These PDFs form the input to our emulator, which interpolates lensing statistics across a broad range of cosmologies without the need for additional costly simulations.

It is important to mention that all of our calculations are performed in the source plane. The transformation of the simulation outputs from the lens plane to the source plane is carried out by scaling with the magnification weight \citep{Takahashi:2011qd}. This is necessary because the results in the source plane differ from those in the lens plane due to magnification or de-magnification effects caused by the matter content of the intervening region. As a sanity check, we verified at the map level, before PDF construction, that:
\begin{align} \label{mu1}
\langle \mu \rangle_{\text{source}} = \frac{\sum_i \mu_i \cdot \left(1/\mu_i\right)}{\sum_i \left(1/\mu_i\right)} = \frac{N}{\sum_i (1/\mu_i)} \approx 1 \,.
\end{align}

\subsection{Lensing map resolution}

Finally, we determine the optimal grid size $r_{\text{grid}}$ to effectively exploit the nonlinear structures probed by the simulations without entering the shot-noise dominated regime. Following the approach of \citet{Takahashi:2011qd}, we model the convergence variance as:
\begin{align}
\langle \kappa^2 \rangle = \frac{9}{8\pi} H_0^4 \Omega_m^2 \int_{0}^{z_s} \frac{\d z}{H(z)} &(1+z)^2 W^2(z_l, z_s) \label{eq:var_cov} \\
&\times \int \d k \, k \left[ P(k,z) + \frac{1}{n} \right] e^{-(k/k_{\text{cut}})^2} , \nonumber
\end{align}
where $P(k,z)$ is the nonlinear matter power spectrum obtained from \texttt{CAMB} (HaloFit, \texttt{mead2020} version).\footnote{\url{https://camb.readthedocs.io}}
The shot-noise contribution is modeled by the term $1/n$, where $n$ denotes the particle number density in the simulations. The Gaussian smoothing introduced by the finite grid size \( r_{\rm grid} \) is modeled by the exponential factor $e^{-(k/k_{\text{cut}})^2}$, where the cutoff wavenumber is defined as $k_{\text{cut}} = 2\pi \alpha / r_{\rm grid}$. We adopt $\alpha = 0.2$, following \citet{Takahashi:2011qd}, which provides an empirical mapping between the smoothing effects of a cubic top-hat filter and a Gaussian kernel.

\begin{figure}
\centering 
\includegraphics[width= \columnwidth]{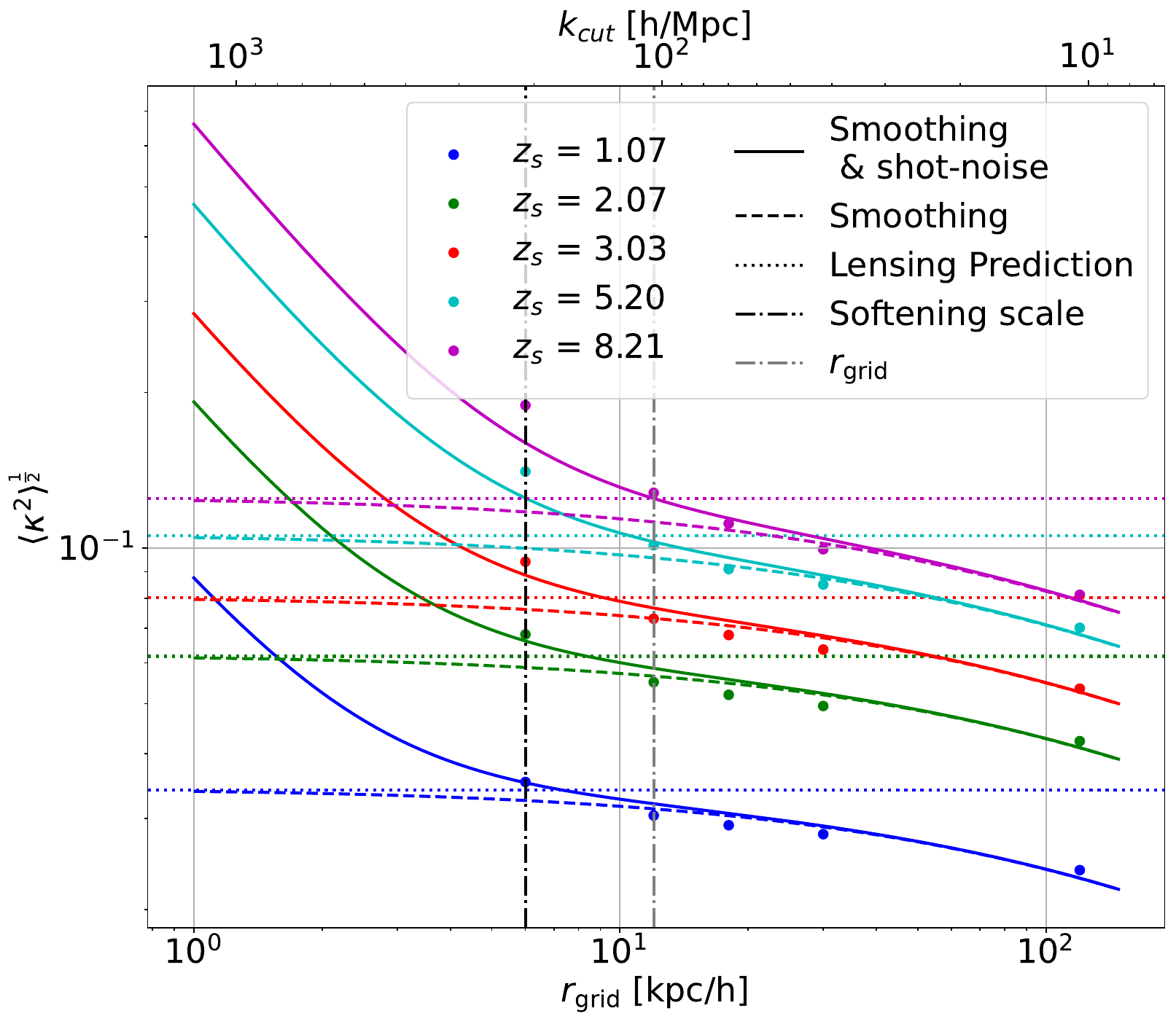}
\caption{Convergence root mean square as a function of grid size, used to determine the optimal resolution before entering the shot-noise dominated regime. The chosen optimal value (12 kpc/h) balances noise and resolution.}
\label{fig:cov_var}
\end{figure}

From Figure~\ref{fig:cov_var}, we conclude that $r_{\rm grid}=12\,\mathrm{kpc}\,h^{-1}$ is an optimal grid size, as it marks the transition to the shot-noise-dominated regime. The convergence test shows that reducing $r_{\rm grid}$ below this value mainly amplifies particle shot noise, rather than adding physically meaningful small-scale information. As a result, magnification PDFs computed at smaller $r_{\rm grid}$ become increasingly dominated by numerical artifacts. Any additional sensitivity to unresolved structure is expected to affect mainly the extreme high-magnification tail. Moreover, baryonic effects dominate the PDF at $\mu \gtrsim 3$ \citep{Castro:2017tbn}, and since baryons are not included in our simulations, exploring smaller grid sizes is beyond the scope of this first work. For $r_{\rm grid}=12\,\mathrm{kpc}\,h^{-1}$, the resulting two-dimensional density, convergence, and shear maps have a resolution of $20833\times 20833$ pixels.

We also stress that sources are treated as pointlike. In this framework, $r_{\rm grid}$ sets the comoving pixel size used to project the density field, and thus the effective angular resolution of the lensing maps probed by a point-like beam. The emulator should therefore be interpreted as predicting an effective, resolution-limited point-source magnification PDF. Sub-grid small-scale structure is not captured by our simulations and is therefore not included in the present modeling; for discussions and possible prescriptions to account for effects such as microlensing by compact objects, we refer to \citet{Bosca:2022viy} and references therein. While baryonic and sub-grid effects can be important in the extreme tail, they are not expected to substantially affect the bulk of the PDF, which is the primary target of the present emulation.

\subsection{Dataset generation and availability} \label{dataset}

At this stage, we produced a dataset consisting of approximately 2000 lensing PDFs spanning 70 cosmological models, publicly available at \href{https://github.com/Turkero/ACE-Lensing}{github.com/Turkero/ACE-Lensing}. 
The lensing PDFs were constructed by binning the magnification maps into 5000 logarithmic bins over the range $0.1 < \mu < 6$.

The choice of the maximum magnification is somewhat arbitrary. In tests, setting a value greater than 6 degraded the performance of the PCA dimensionality reduction because the high-$\mu$ tail disproportionately weighted the variance, downweighting the body of the PDF. 
Conversely, adopting a smaller $\mu_{\max}$ prematurely truncated the high-magnification tail relative to the low-magnification wing. 
Although modeling is less reliable at very high magnifications, we retain support up to $\mu=6$ to ensure a sufficiently wide domain. 
For supernova and gravitational-wave applications, the residual inaccuracy near the upper bound is expected to introduce at most a small bias.

\section{PDF feature extraction}
\label{sec:features}

\subsection{Standardization of the PDFs}

We adopt Principal Component Analysis (PCA) to efficiently represent the PDFs. 
Before applying PCA, we standardize the PDFs to place them on a comparable scale and shape, which improves the performance of the PCA decomposition and, consequently, the training of our models.
We standardized the PDFs as follows:
\begin{gather}
    \mu_{\rm std} = \frac{\mu - \bar{\mu}}{\sigma} \,, \qquad
    P(\mu_{\rm std}) = P(\mu) \cdot \sigma \label{eq:std1} \,.
\end{gather}
Here, $ \mu $ denotes the lensing magnification, with mean $ \bar{\mu} $ and standard deviation $ \sigma $. The variable $ \mu_{\rm std} $ is the standardized magnification, and $ P(\mu_{\rm std}) $ is its corresponding probability distribution function.

After standardization, we trimmed the PDF tails for values smaller than $\max({\rm PDF})/10^4$ to remove noisy artifacts. This operation does not affect the PDF region relevant for emulation. We then interpolated all PDFs onto a common magnification grid of 5000 evenly spaced points, spanning from $-2$ to $8$ standard deviations ($\sigma_{\rm std}=1$). 
This procedure ensures a fixed, shared grid across all PDFs, eliminating the need for explicit emulation of the grid itself. 
While no PDF extends below $-2$ standard deviations due to intrinsic skewness, some may exhibit tails beyond $8$ standard deviations. However, considering the intended applications of our emulator, the finite resolution of the simulations, and the absence of baryonic physics, extending the standardized PDFs beyond $8\sigma$ would not provide additional meaningful cosmological information.

\subsection{Dimensionality reduction with PCA}
\label{npca_pre}

\begin{figure}
\centering 
\includegraphics[width= \columnwidth]{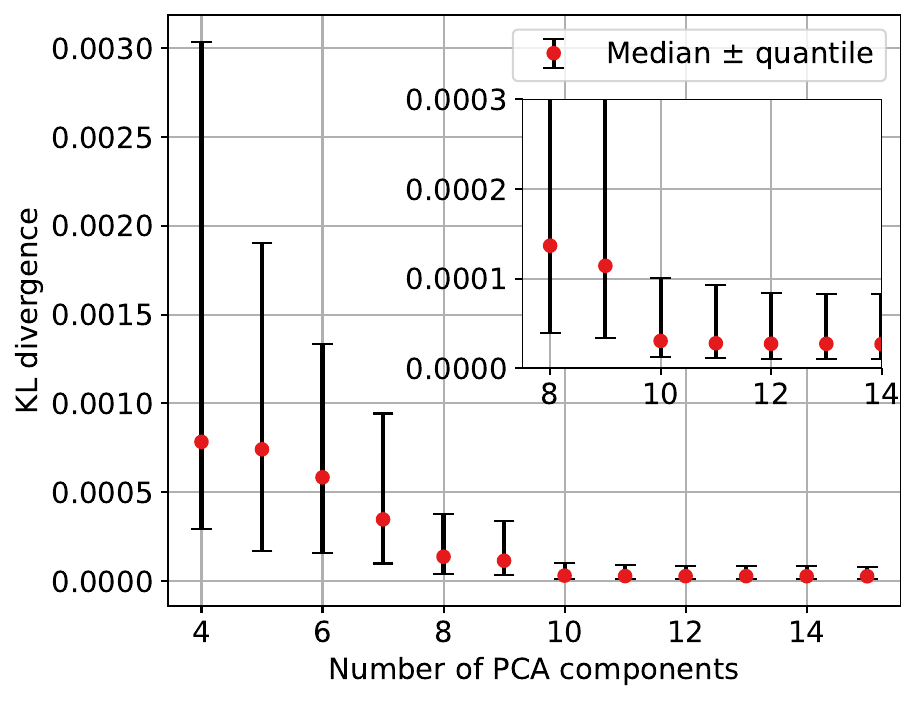}
\caption{KL divergence between original and reconstructed PDFs as a function of the number of PCA components. Error bars indicate the 16th–84th percentile range across 1447 PDFs.}
\label{fig:PCA_comp}
\end{figure}

After standardizing the PDFs, we performed several tests to determine the optimal number of PCA components. The first test involved computing the Kullback–Leibler (KL) divergence between the original and reconstructed PDFs, evaluated over the 1447 PDFs corresponding to redshifts $z < 6$, as a function of the number of PCA components. The KL divergence is defined as:
\begin{gather}
    D_{\text{KL}}(P\parallel Q) = \sum P(x) \log \left( \frac{P(x)}{Q(x)} \right) \,,
\end{gather}
where $P$ is the original PDF and $Q$ is the PCA-reconstructed PDF.
Figure~\ref{fig:PCA_comp} shows that reconstruction fidelity improves with the number of components, reaching a plateau around 10 components, beyond which no significant gain is observed.

\begin{figure}
\centering 
\includegraphics[width= \columnwidth]{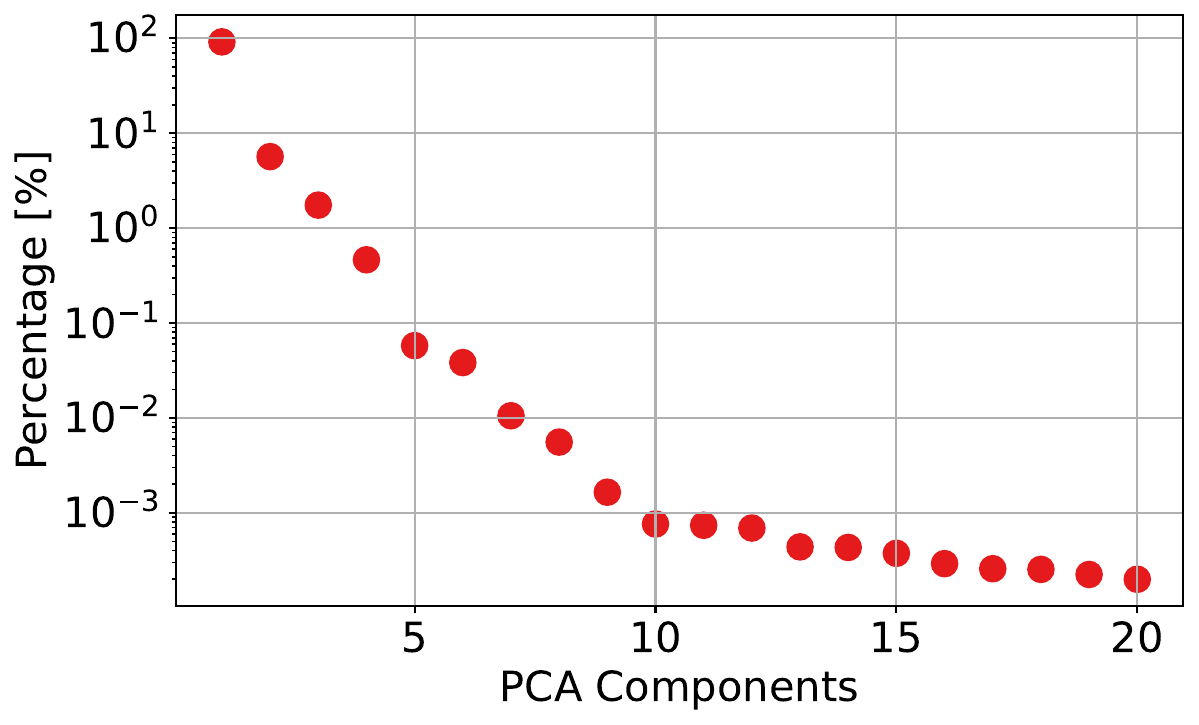}
\caption{Non-cumulative explained variance ratio for each PCA component. The first component captures over 90\% of the total variance.}
\label{fig:variance_exp}
\end{figure}

The relative importance of each PCA component is shown in Figure~\ref{fig:variance_exp}, where the first component alone interestingly accounts for approximately 91\% of the total variance. Sharp drops in explained variance are observed after the 4th, 6th, and 8th components. Once again, we notice a plateau after the 10th component, indicating diminishing returns beyond this point. These thresholds provide natural cutoffs for retaining the most informative features of the PDFs while minimizing redundancy and the risk of overfitting.

\begin{figure}
\centering 
\includegraphics[width= \columnwidth]{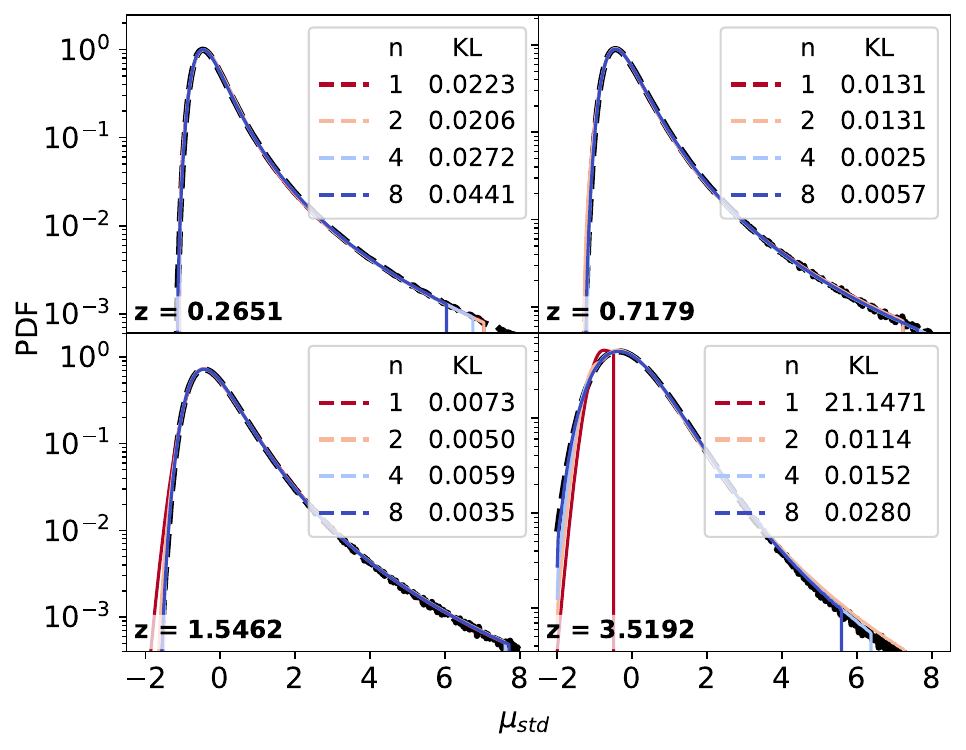}
\caption{Visual comparison of  randomly selected standardized PDFs at representative redshifts. The black line in the background represents the original PDF. The legend shows the KL divergence values corresponding to the PDFs reconstructed with a given number of PCA components.}
\label{fig:pca-ex}
\end{figure}

Figure~\ref{fig:pca-ex} provides a visual comparison of four randomly selected standardized PDFs at representative redshifts. We note that one PCA component indeed captures the main features of each PDF, broadly reproducing their overall shape, and that it is hard to spot by eye the differences after two PCA components in all redshifts. Additional components are nevertheless required to achieve an accurate reconstruction.

\begin{figure*}
\centering
\includegraphics[trim={0 0 0 9.18cm}, clip, width= \textwidth]{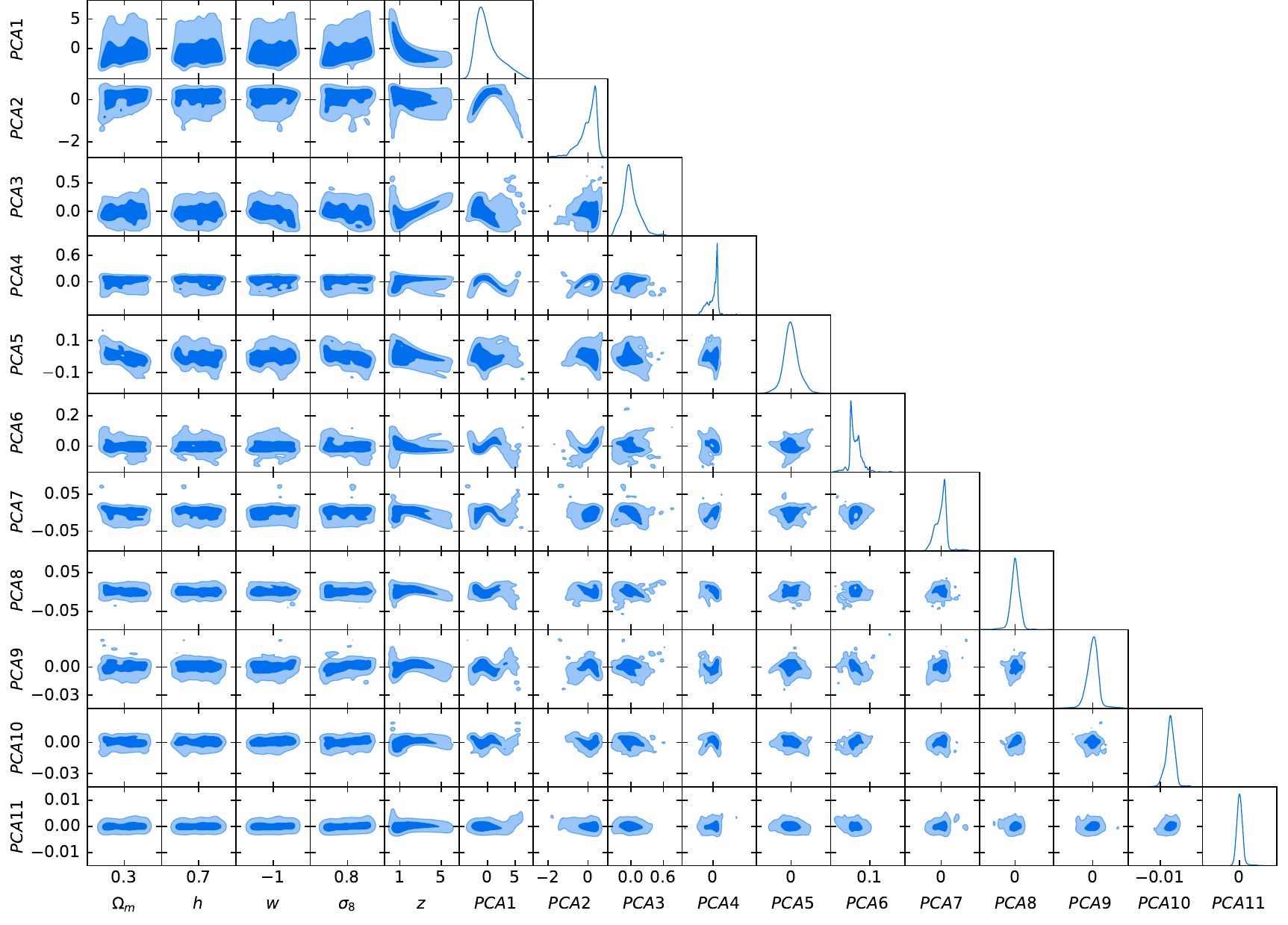}
\caption{Triangular correlation matrix between PCA components and input parameters, summarizing the relationships between the principal modes and the cosmological parameters. Numerical values indicate the nonlinear Spearman correlation coefficients.}
\label{fig:correlation_PCA_det}
\end{figure*}

Finally, we analyzed the correlation between the PCA components and the cosmological parameters, including redshift. 
Figure~\ref{fig:correlation_PCA_det} shows that the first few PCA components display strong non-linear correlations with the input features and among themselves, indicating that they encode meaningful cosmological information relevant for reconstructing the PDFs. As expected, the correlation strength progressively decreases for higher-order components and becomes negligible beyond the tenth, consistent with the trends observed in Figures~\ref{fig:PCA_comp} and~\ref{fig:variance_exp}.

The discussion above focused solely on determining the optimal number of PCA components based on the reconstruction accuracy of the PDFs, without accounting for the additional noise introduced by the machine learning model. As we will show in the next section, it is this model-induced noise that ultimately determines the truly optimal number of components.

\section{Machine learning model}
\label{sec:ML}

As discussed in the previous section, we model the lensing PDF via its mean and standard deviation (whose information is absent in the standardized PDF used for the PCA analysis), together with the first $n$ PCA components.
Although the mean is very close to the theoretical value of unity, we found that to improve numerical accuracy it is useful to model it along side the variance. We remind that the emulator outputs the lensing PDF in the source plane.

\subsection{XGBoost}

To complete the emulator, we require a model that maps the five input features--the cosmological parameters ($\Omega_{\rm m}$, $\sigma_8$, $w$, $h$) and redshift--to the $n+2$ output quantities describing the PDF. 
Given the size of the training set, we employ \texttt{XGBoost}, which builds an ensemble of gradient-boosted decision trees to model nonlinear relations efficiently. 
While neural networks may outperform tree-based methods on substantially larger datasets, they are unlikely to provide significant advantages in our setting.

Consequently, we build $n+2$ \texttt{XGBoost} models and perform an individual hyperparameter search for each of them over a parameter grid to optimize performance. The grid includes variations in the number of estimators, learning rate, tree depth, and sampling ratios, as summarized in Table~\ref{tab:hp}.
We split our set of 1447 PDFs into 80\% for training (1157 PDFs) and 20\% for testing (290 PDFs).

\begin{table}
\centering
\caption{Hyperparameter search grid for the \texttt{XGBoost} model.}
\label{tab:hp}
\begin{tabular}{ll}
\toprule
Parameter & Values \\
\midrule
\texttt{n\_estimators} & 50, 100, 200 \\
\texttt{learning\_rate} & 0.01, 0.1, 0.2 \\
\texttt{max\_depth} & 3, 5, 7 \\
\texttt{subsample} & 0.7, 1.0 \\
\texttt{colsample\_bytree} & 0.7, 1.0 \\
\bottomrule
\end{tabular}
\end{table}

\subsection{Choosing the optimal number of PCA components}

\begin{figure}
\centering
\includegraphics[width=\columnwidth]{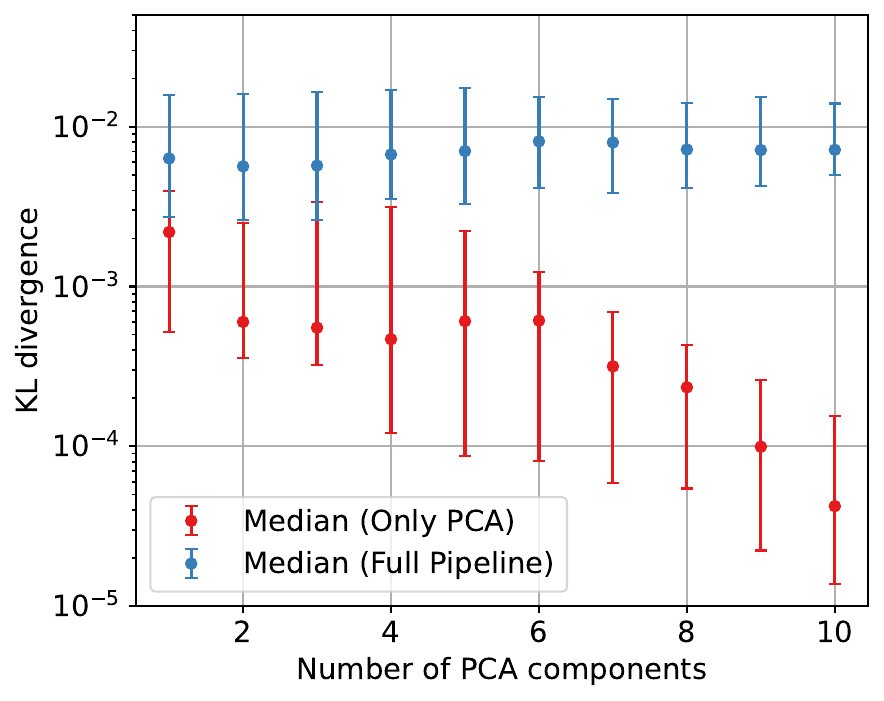}
\caption{Median KL divergence as a function of the number of PCA components. Error bars indicate the 16th–84th percentile range across the test set (290 PDFs). Red points correspond to an analysis similar to that of Figure~\ref{fig:PCA_comp}, accounting only for information loss due to dimensionality reduction. Blue points include the full pipeline, incorporating ML prediction errors.}
\label{fig:KL_div_comparison}
\end{figure}

In Section~\ref{npca_pre}, we concluded that up to 10 PCA components were sufficient to accurately reconstruct the lensing PDF. However, that analysis considered only the PCA reconstruction step, ignoring the additional prediction noise introduced by the ML modeling stage. Here, we refine this analysis by accounting explicitly for the full pipeline, including the ML-induced uncertainty. We expect the conclusions of this Section to evolve as the training set expands and ML techniques improve.

Figure~\ref{fig:KL_div_comparison} compares the KL divergence as a function of the number of PCA components, contrasting the scenario that includes only PCA dimensionality reduction with the full pipeline of Figure~\ref{fig:pipeline}, which incorporates ML prediction errors.
We ultimately selected 4 PCA components, as this choice provides an optimal balance between reconstruction accuracy and model stability, yielding greater robustness against outliers. 
Indeed, Figure~\ref{fig:KL_div_comparison} presents median KL divergence values, which are robust measures of central tendency but do not fully capture the spread or presence of outliers. 
We observed that the number of outliers decreases when using 4 PCA components.

\subsection{PDF reconstruction and post-processing}

Using the $n$ PCA components predicted by XGBoost, we first reconstruct the standardized PDF by applying the inverse PCA transformation.
Artifacts occasionally appear in the reconstructed PDFs and so we apply a systematic post-processing step. Since the magnification PDFs are expected to be monotonic on both sides of the mode, we search for local maxima and minima, and then smooth these anomalies to restore monotonicity.
Next, we apply a Fourier transformation and filter out high-frequency components to remove residual noise.

To recover the original (unstandardized) PDF, we apply the inverse transformation using the predicted values for the mean and standard deviation:
\begin{gather}
    \mu = \sigma_{\rm pred} \, \mu_{\rm std} + \bar{\mu}_{\rm pred} \,, \qquad
    P(\mu) = \frac{1}{\sigma_{\rm pred}} \, P_{\rm pred}(\mu_{\rm std}) \label{eq:std2} \,,
\end{gather}
where $P_{\rm pred}(\mu_{\rm std})$ is the standardized PDF reconstructed from the predicted PCA components.
We renormalize the PDF to preserve unit integral.

Due to flux conservation in the source plane, the mean of the magnification PDF is expected to be unity, see Eq.~\eqref{mu1}. To exactly enforce this, we rescale the emulator output as follows. We define the rescaled magnification and corresponding PDF by
\begin{gather} \label{mu1biju}
\mu' = \frac{\mu}{m_1} \,, \qquad
P'(\mu') = m_1 \, P(\mu) \,, \qquad
m_1 = \int \mu \, P(\mu) \, d\mu \,,
\end{gather}
which ensures that
\begin{gather}
\int P'(\mu') \, d\mu' = 1 \,, \qquad
\int \mu' \, P'(\mu') \, d\mu' = 1 \,.
\end{gather}
Although this correction is very small, it ensures that the theoretical mean is recovered, yielding an unbiased measurement of the luminosity distances.

\subsection{ML performance}

\begin{figure}
\centering 
\includegraphics[width=\columnwidth]{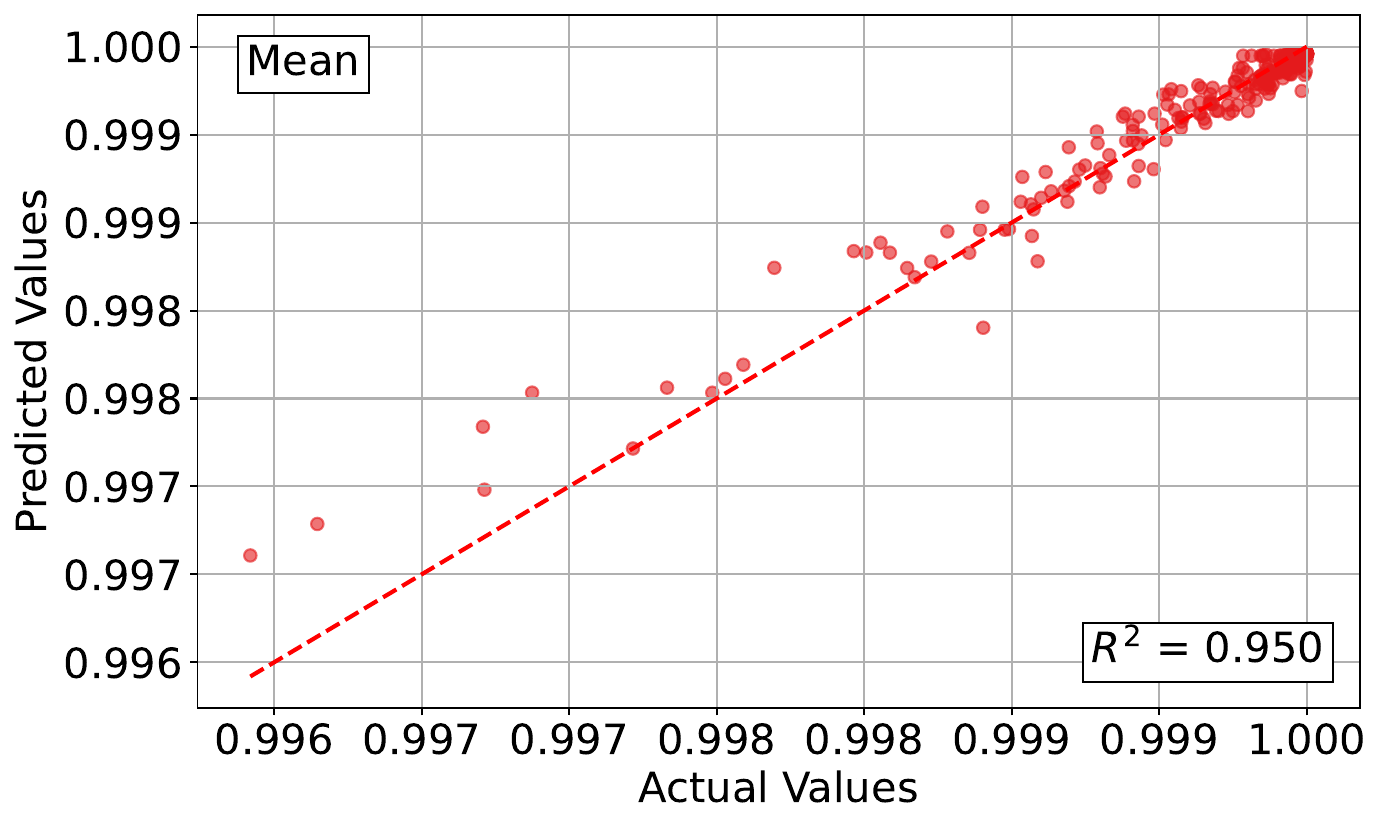}
\caption{Performance of the XGBoost regressor in predicting the mean of the magnification PDFs.}
\label{fig:mean}
\end{figure}

As discussed earlier in Eq.~\eqref{mu1}, our maps yield a mean value very close to the theoretical expectation of unity. However, as mentioned in Section~\ref{dataset}, we bin the PDF within the magnification range $0.1 < \mu < 6$, thereby excluding highly magnified events for which our physical model is unreliable and that are not relevant for the emulator’s purpose. This truncation leads to a mean magnification slightly below unity, since only the high-magnification tail is systematically removed. The resulting bias, however, is well correlated with redshift and cosmology, as evidenced by the good performance of the XGBoost model in reconstructing the mean (Figure~\ref{fig:mean}), allowing for an accurate modeling and \textit{a posteriori} correction of the PDF mean, as implemented through Eq.~\eqref{mu1biju}.

\begin{figure}
\centering 
\includegraphics[width= \columnwidth]{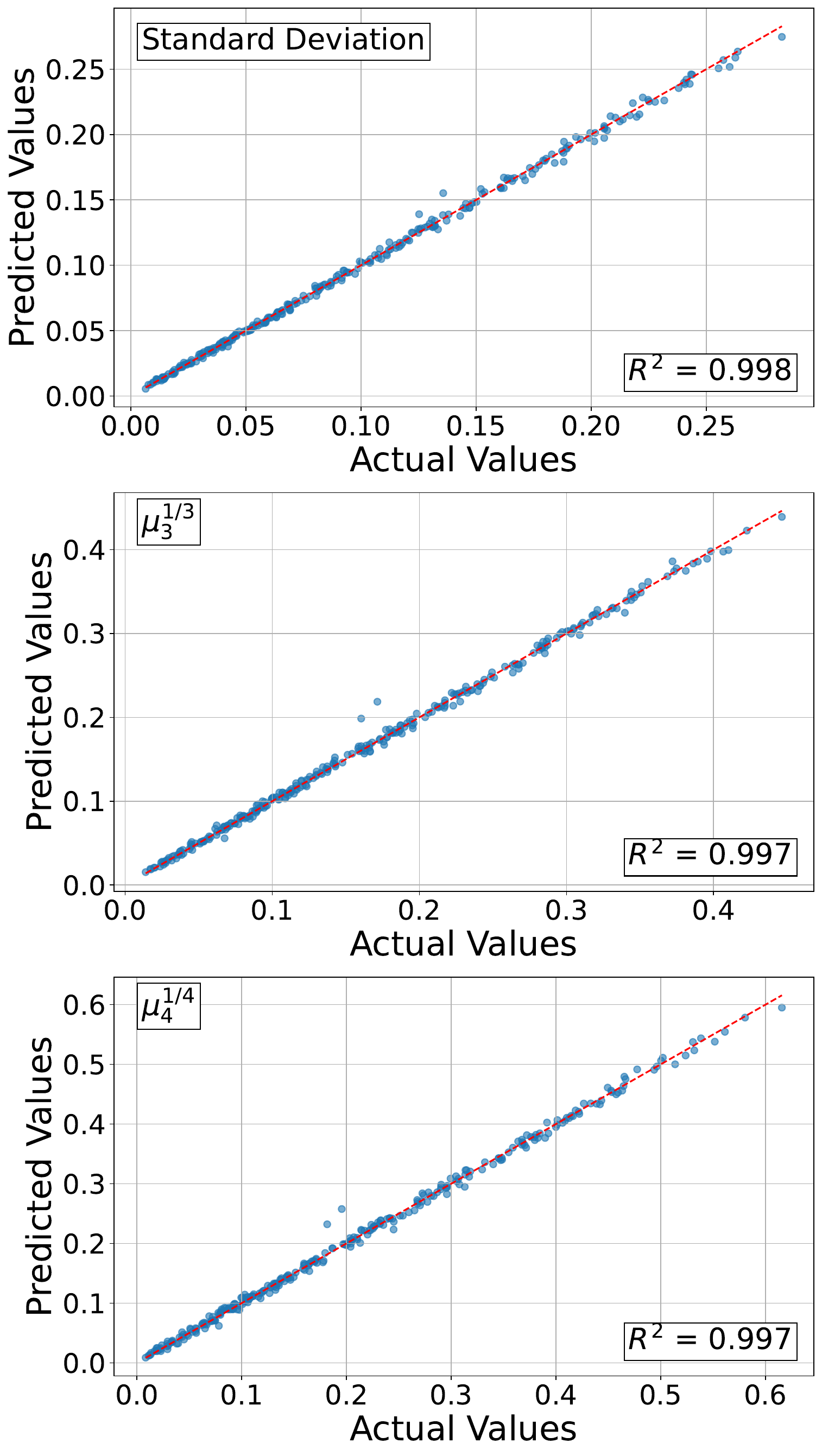}
\caption{Performance of the XGBoost regressor in predicting the second-to-fourth moments of the magnification PDFs.}
\label{fig:moments}
\end{figure}

Figure~\ref{fig:moments} illustrates the excellent performance of the XGBoost regressor in predicting the second-to-fourth moments of the magnification PDFs, whose accurate modeling is crucial for extracting cosmological information from lensing scatter via the method of the moments \citep[MeMo,][]{Quartin:2013moa}.

\begin{figure}
\centering 
\includegraphics[width= \columnwidth]{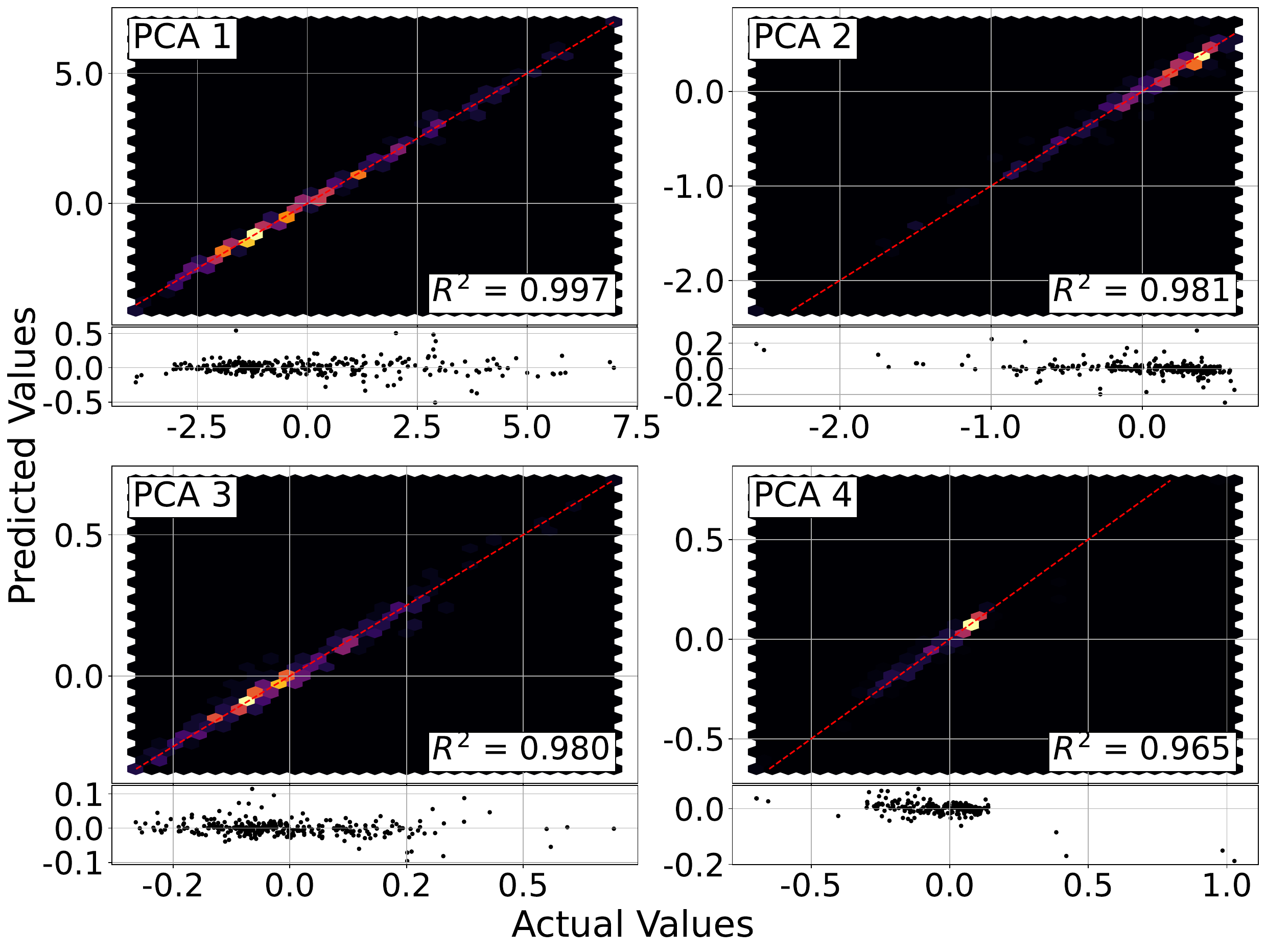}
\caption{Performance of the XGBoost regressor in predicting the PCA component values.}
\label{fig:PCA8}
\end{figure}

Finally, Figure~\ref{fig:PCA8} shows the accuracy in predicting the first 4 PCA component values.
The accuracy in predicting the $n$-th PCA component decreases with increasing $n$, as higher-order components typically capture subtler and noisier variations.

\section{Emulator performance}
\label{sec:perf}

\begin{figure}
\centering 
\includegraphics[width=\columnwidth]{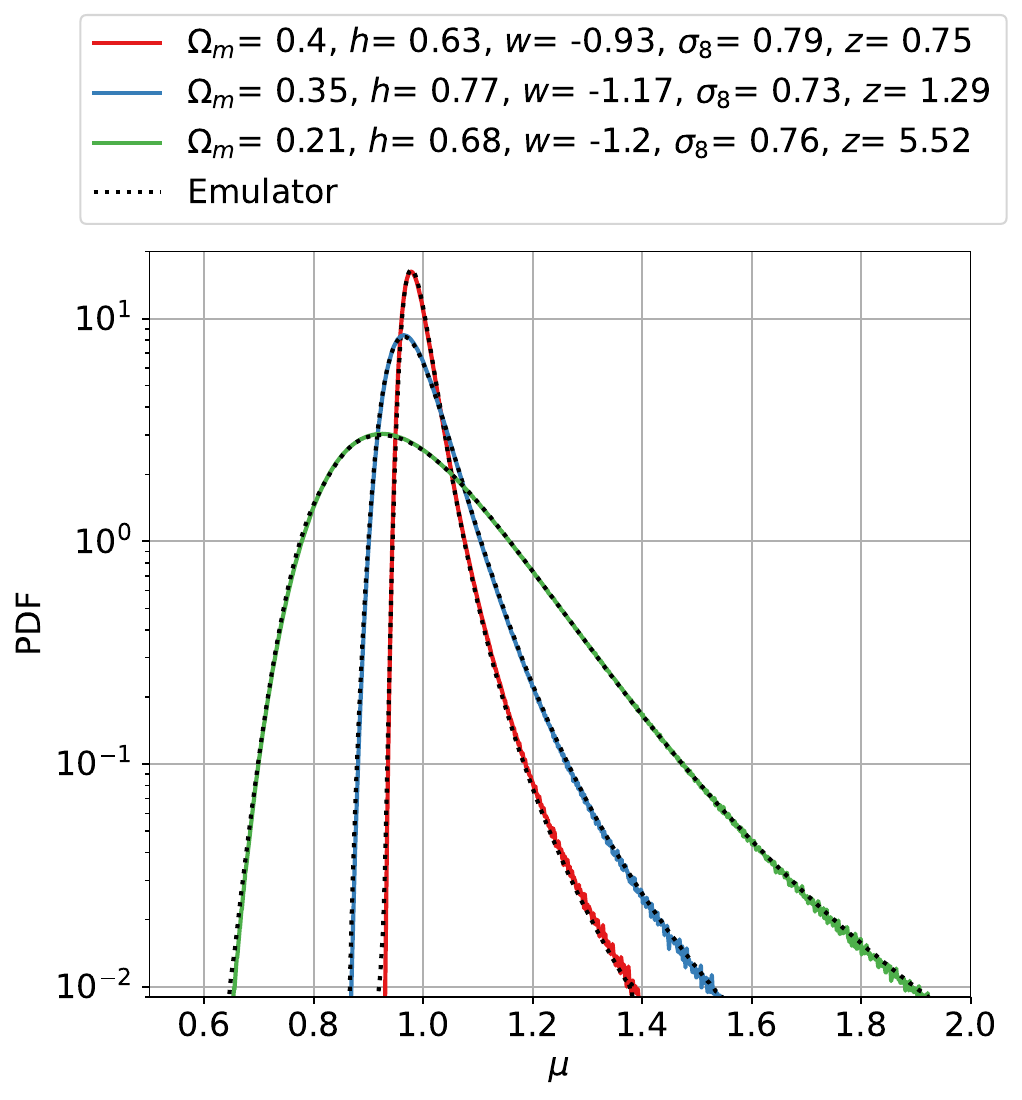}
\caption{
Comparison between emulated PDFs (dotted lines) and actual PDFs from the test set (solid lines) at representative redshifts.}
\label{fig:emulator}
\end{figure}

Figure~\ref{fig:emulator} visually compares selected emulated PDFs (solid lines) with the actual PDFs from the test set (dotted lines) at four representative redshifts. The plots confirm the high quality of the emulation, consistent with the low KL divergence values reported in the legend.

\begin{figure}
\centering 
\includegraphics[width=\columnwidth]{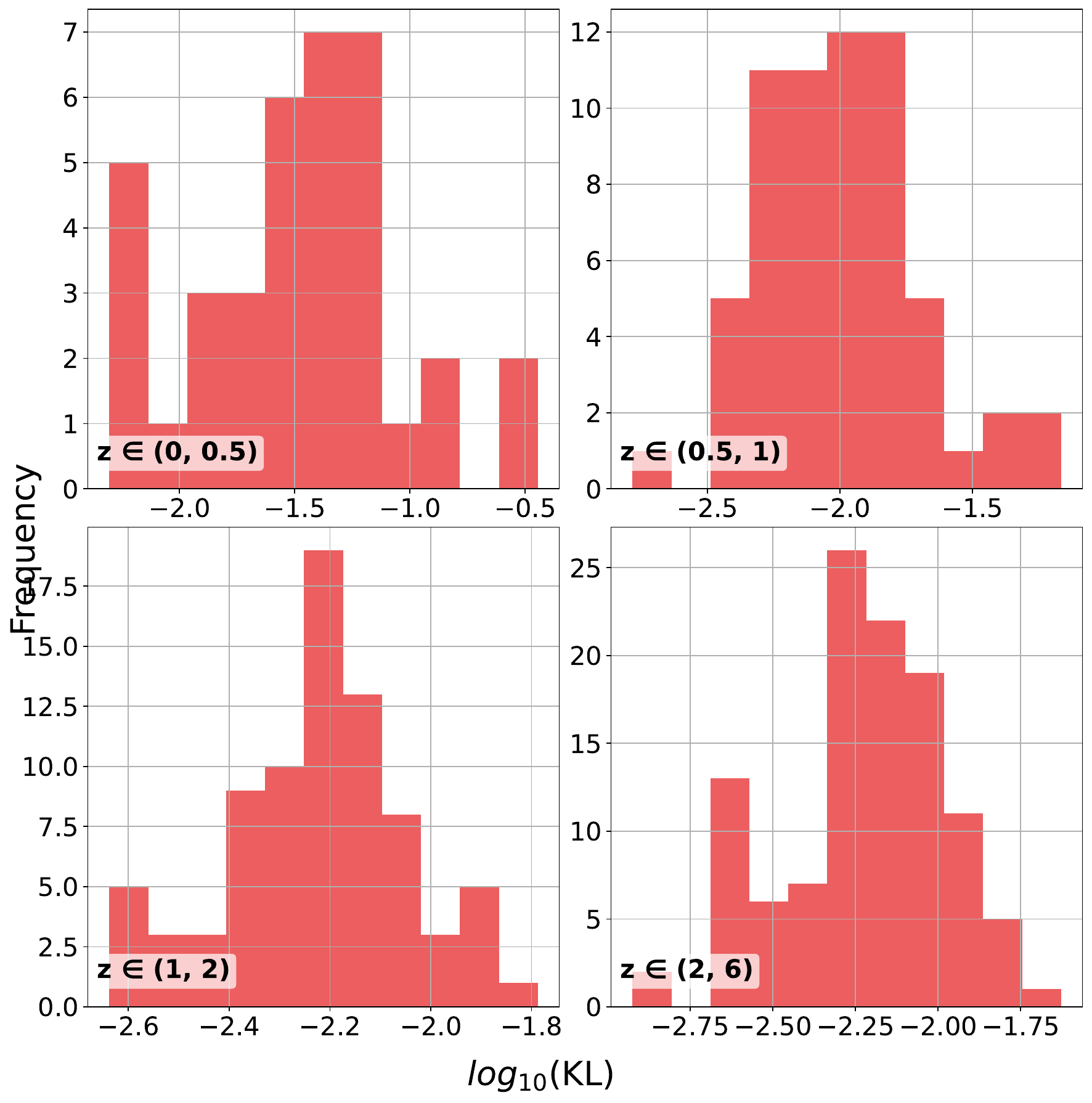}
\caption{Histogram of KL divergence values over the test set (290 PDFs) across the full pipeline using 4 PCA components.}
\label{fig:KL}
\end{figure}

Figure~\ref{fig:KL} shows the histogram of KL divergence values between the reconstructed and original PDFs over the test set, after applying the full pipeline with 4 PCA components. The median KL divergence is 0.0069, with nearly all PDFs exhibiting values below 0.1. 
The four redshift bins were defined according to the following rationale. 
The first bin, $[0, 0.5]$, includes sources for which lensing has a negligible impact. 
The second bin, $[0.5, 1]$, corresponds to the redshift range most relevant for current Type Ia supernova samples, such as DESY5 \citep{DES:2024jxu} and Pantheon+ \citep{Scolnic:2021amr}. 
The third bin, $[1, 2]$, will be important for the supernovae to be discovered by the Nancy Grace Roman Space Telescope \citep{Rose:2021nzt}. 
Finally, the last bin, $[2, 6]$, will be particularly relevant for the high-redshift gravitational-wave events \citep{Menote:2025zmn} detected by third-generation observatories, such as the Einstein Telescope \citep{ET:2019dnz}.

\begin{figure*}
\centering 
\includegraphics[width=\textwidth]{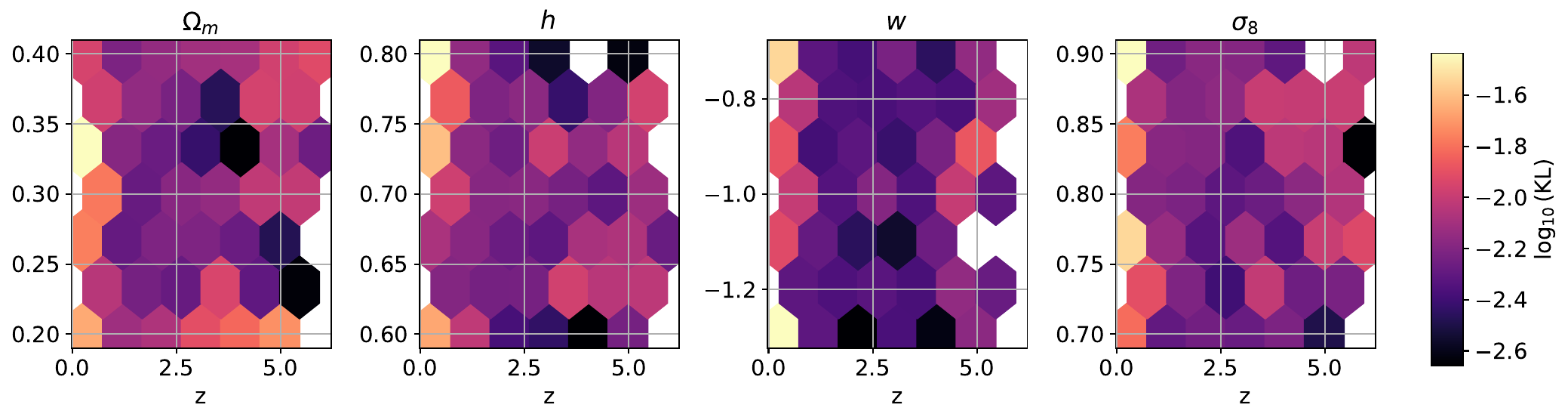}
\caption{KL divergence values as functions of cosmological parameters and redshift, for the full pipeline with 4 PCA components.}\label{fig:hexa}
\end{figure*}

Finally, Figure~\ref{fig:hexa} presents the KL divergence as a function of cosmological parameters. No clear patterns are observed, indicating that the performance is primarily limited by noise in the ML predictions, which could be reduced with a larger training set. A mild trend of increasing KL divergence with higher $\sigma_8$ and redshift is visible, though it remains noisy.

\section{Conclusions} \label{conclusions}

In this work, we presented a comprehensive framework to accurately model the gravitational lensing magnification probability distribution function (PDF) using cosmological $N$-body simulations. Our approach addresses the limitations of previous analytic and semi-analytic approximations, particularly their inadequacies in capturing the nonlinear structures crucial for accurate cosmological inference.

We constructed a robust pipeline encompassing numerical simulations, past light cone generation, convergence and shear map computation, dimensionality reduction via PCA, and machine learning emulation employing the XGBoost algorithm. This pipeline enables efficient interpolation of lensing PDFs across a broad parameter space, covering cosmological parameters ($\Omega_{\rm m}$, $\sigma_8$, $w$, $h$) and source redshifts ($0.2 \leq z \leq 6$).

We  validated our methodology, demonstrating good agreement between the \texttt{SLICER} and \texttt{PAINLESS} approaches and identifying an optimal spatial resolution ($r_{\rm grid}=12$~kpc$h^{-1}$) to mitigate shot noise while preserving nonlinear structure information. By performing dimensionality reduction using PCA, we found that employing four PCA components provides a robust balance between accuracy and stability, effectively capturing the essential features of the magnification PDFs.

Our machine-learning emulator, \texttt{ace\_lensing}, achieves high accuracy, with a median KL divergence of 0.007 between emulated and original PDFs. 
Validation on the test set confirmed that the model reliably reproduces the detailed shapes and statistical properties of the PDFs across the explored parameter range, showing no significant degradation for specific parameter combinations or redshifts.

The emulator \href{https://github.com/Turkero/ACE-Lensing}{\texttt{ace\_lensing}} is publicly available, providing the cosmology and astrophysics communities with an efficient and precise tool for extracting cosmological information from gravitational lensing magnification data.
The Python emulator module \texttt{ace\_lensing} provides both the full lensing magnification PDF and standalone functions to directly predict its second-to-fourth moments.

In future work, we will extend the emulator on the modeling side by enlarging the training set with additional small-volume simulations to improve the stability of the ML predictions, a limited number of large-volume boxes to assess the impact of cosmic variance, and a representative set of hydrodynamical simulations to quantify and calibrate baryonic effects. This effort aims at reducing both statistical and systematic sources of bias in the emulator.

\begin{acknowledgements}
TT acknowledges financial support from FAPES (Brazil), CAPES (Brazil) and CNPq (Brazil). VM acknowledges partial support from CNPq (Brazil) and FAPES (Brazil). MQ is supported by FAPERJ (Brazil) project E-26/201.237/2022, CNPq (Brazil) and CAPES (Brazil). 
SB acknowledges support from the "Fondazione ICSC" National Recovery and Resilience Plan (PNRR) Project ID CN-00000013 "Italian Research Center on High-Performance Computing, Big Data and Quantum Computing" funded by MUR Missione 4, Componente 2 Investimento 1.4: "Potenziamento strutture di ricerca e creazione di "campioni nazionali di R$\&$S (M4C2-19 )" - Next Generation EU (NGEU). 
We  acknowledge the use of the HOTCAT computing infrastructure of the Astronomical Observatory of Trieste of the National Institute for Astrophysics (INAF, Italy) \citep[see][]{2020ASPC..527..303B,2020ASPC..527..307T}, of the computing center  of Cineca, of the Santos Dumont supercomputer of the National Laboratory of Scientific Computing (LNCC, Brazil), of the \href{https://computacaocientifica.ufes.br/scicom}{Sci-Com Lab} of the Department of Physics at UFES, supported by FAPES, CAPES, and CNPq 
and of the computational resources of the joint CHE / Milliways cluster, supported by a FAPERJ grant E-26/210.130/2023.
\end{acknowledgements}

\bibliographystyle{aaArxivDoi}
\bibliography{biblio}

\begin{appendix}


\section{Pipeline}
\label{ap:pipeline}

\begin{figure}
\centering 
\includegraphics[width= \columnwidth]{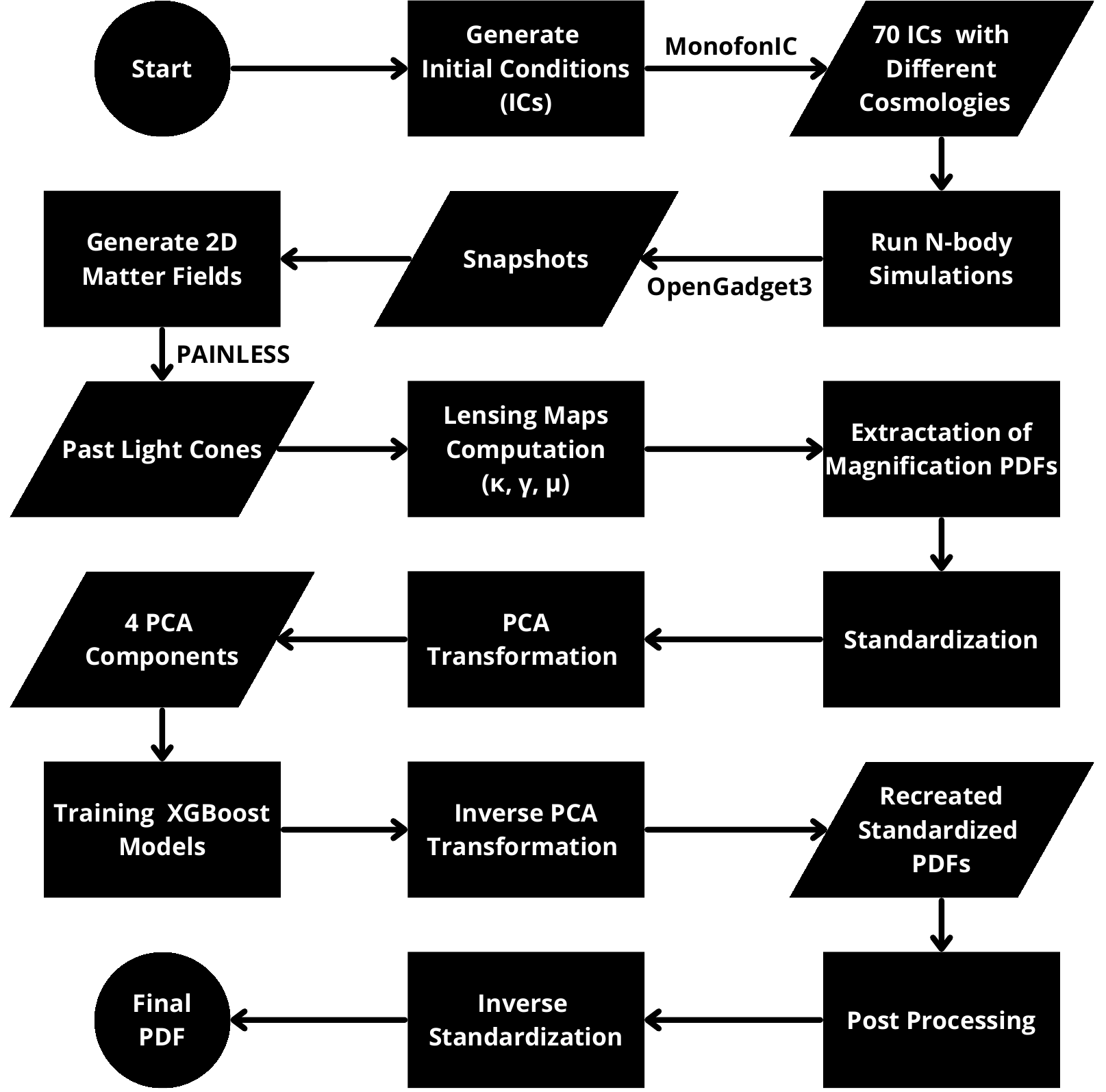}
\caption{Full pipeline of the lensing PDF emulator.}
\label{fig:pipeline}
\end{figure}

Figure~\ref{fig:pipeline} summarizes the pipeline adopted to build the \texttt{ace\_lensing} emulator.

\section{\texttt{SLICER} vs \texttt{PAINLESS}}
\label{ap:slicingpain}

Here we compare the \texttt{SLICER} and \texttt{PAINLESS} methods. 
To validate and quantify potential discrepancies between them, we define the effective grid size $r_{\text{grid}}^{\text{eff}}$ for \texttt{PAINLESS} that corresponds to a given angular resolution $\theta_{\text{grid}}$ used in \texttt{SLICER} \citep[see][]{Castro:2017tbn}:
\begin{gather}\label{rgrid}
r_{\text{grid}}^{\text{eff}} =
\frac{\int_{0}^{z_s} \d z \, r_{\text{grid}}(z) \, G^2(z) \, \frac{(1+z)^2}{H(z)} W^2(z, z_s)}
{\int_{0}^{z_s} \d z \, G^2(z) \, \frac{(1+z)^2}{H(z)} W^2(z, z_s)} \,,
\end{gather}
where $r_{\text{grid}}(z) = \theta_{\text{grid}}  \DM(z)$ is the comoving scale associated with the fixed angular resolution $\theta_{\text{grid}}$.

\begin{figure}
\centering 
\includegraphics[width= \columnwidth]{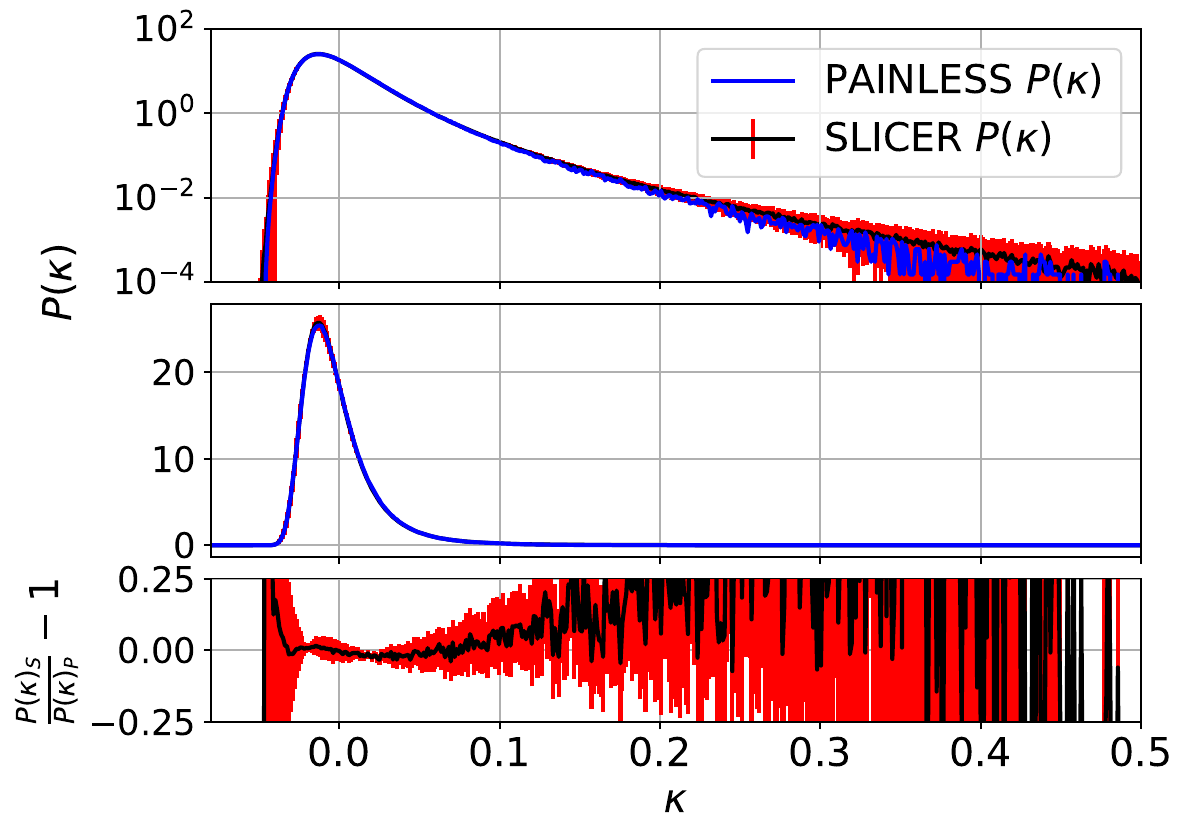}
\caption{Mean of 50 \texttt{SLICER} realizations (with standard deviation) with $\theta_{\text{grid}}=14.06$~arcsec compared to a single \texttt{PAINLESS} realization with $r_{\text{grid}}=106.8$~kpc for a source at $z=1$.}
\label{fig:SLICERvsPAINLESS}
\end{figure}

We compared the convergence PDFs from \texttt{SLICER} with $\theta_{\text{grid}}=14.06$~arcsec to those from \texttt{PAINLESS}, using $r_{\text{grid}}^{\text{eff}}$ from Eq.~\eqref{rgrid}. Each simulation yields a single \texttt{PAINLESS} PDF, since all particles are used. In contrast, \texttt{SLICER} samples only a fraction of particles along discrete PLCs; to fully exploit the box we generated 50 PLC realizations per simulation. Figure~\ref{fig:SLICERvsPAINLESS} compares the mean \texttt{SLICER} PDF to the corresponding \texttt{PAINLESS} PDF: the agreement is excellent for $\kappa \lesssim 0.1$, while systematic deviations appear in the high-$\kappa$ tails, where sampling noise and rare nonlinear structures dominate. The increasing error bars at large $\kappa$ reflect the smaller number of contributing pixels and the sensitivity to line-of-sight selection. Finally, Eq.~\eqref{rgrid} provides only an approximate mapping between the two resolution scales; residual discrepancies in the tails are therefore expected because the two methods probe the density field with different spatial samplings.

\begin{figure}
\centering 
\includegraphics[width= \columnwidth]{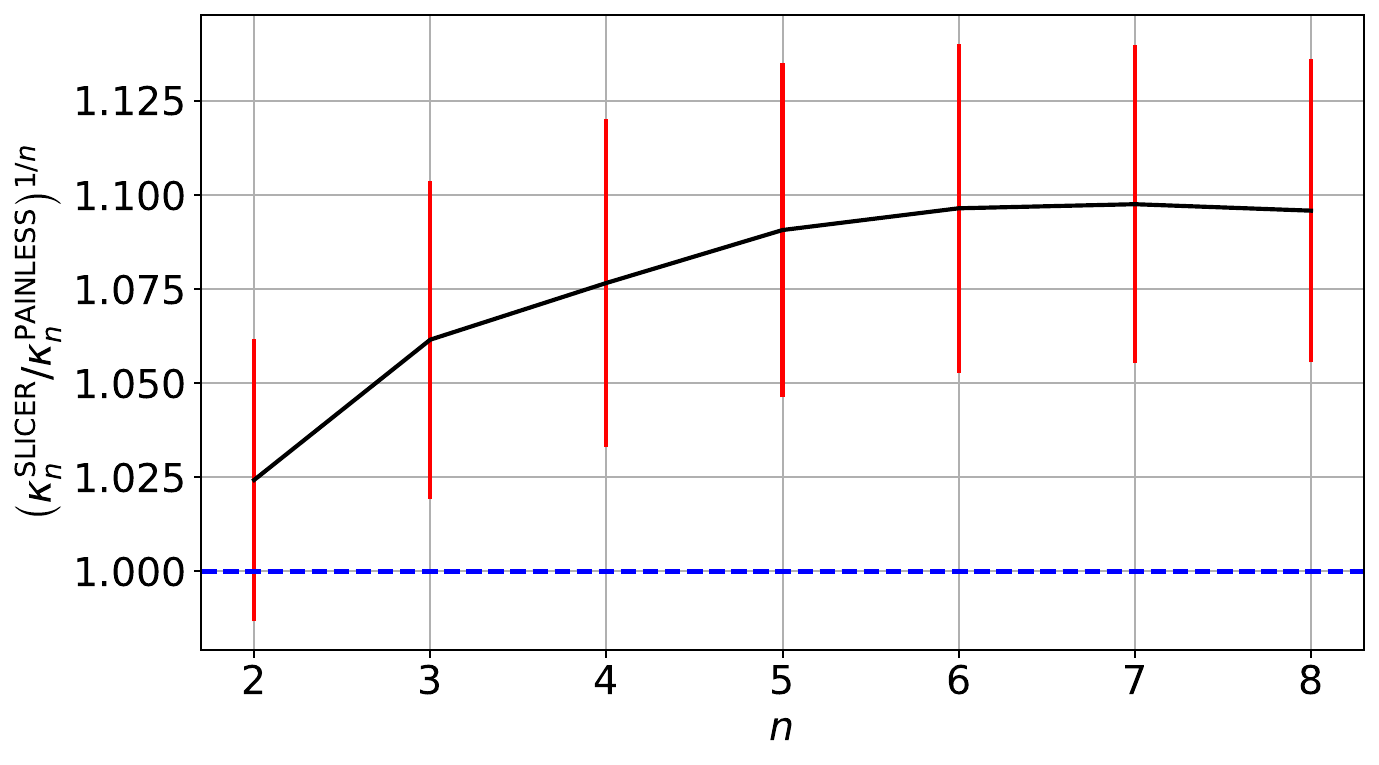}
\caption{Comparison of central moments between mean \texttt{SLICER} PDF (50 realizations) and \texttt{PAINLESS} PDF.}
\label{fig:SLICERvsPAINLESS-mom}
\end{figure}

Next, in Figure~\ref{fig:SLICERvsPAINLESS-mom}, we compare the $n$-th central moments, defined as $\kappa_n = \left\langle (\kappa - \langle \kappa \rangle)^n \right\rangle$. To facilitate this comparison, we plot the ratio $(\kappa^{\rm \texttt{SLICER}}_n / \kappa^{\rm \texttt{PAINLESS}}_n)^{1/n}$. We obtain values close to unity for small $n$, exhibiting modest upward deviations at higher-order moments, consistent with the finding from Figure~\ref{fig:SLICERvsPAINLESS}. 
Overall, these results reinforce the statistical consistency between the \texttt{SLICER} and \texttt{PAINLESS} methods.




\end{appendix}

\end{document}